\documentclass[useAMS,usenatbib]{mnras}

\usepackage{amsmath,amsfonts,graphicx,txfonts,fleqn,color,tabularx,verbatim,fixltx2e,gensymb}
\usepackage{makecell}
\renewcommand*{\vec}[1] {\boldsymbol{#1}}
\newcommand*  {\kms}   {\mbox{km\,s$^{-1}$}}

\title[Debris discs in stellar open clusters]{The fate of planetesimal discs in young open clusters: implications for 1I/'Oumuamua, the Kuiper belt, the Oort cloud and more}
\author[Hands, Dehnen, Gration, Stadel \& Moore]{T.O.Hands$^1$ \thanks{email:
tomhands@physik.uzh.ch}, W. Dehnen$^{2,3}$, A. Gration$^2$, J. Stadel$^1$, B. Moore$^1$ 
\\$^1$Institut f\"ur Computergest\"utzte Wissenschaften, Universit\"at Z\"urich, Winterthurerstrasse 190, 8057 Z\"urich, Switzerland
\\$^2$Department of Physics \& Astronomy, University of Leicester, University Road, Leicester, LE1 7RH, UK
\\$^3$Universit\"ats-Sternwarte der Ludwig-Maximilians-Universit\"at, Scheinerstrasse 1, Mu\"unchen D-81679, Germany
}
\hypersetup{draft}
\begin{document}

\pagerange{\pageref{firstpage}--\pageref{lastpage}} \pubyear{2019}

\maketitle

\label{firstpage}

\begin{abstract}
We perform $N$-body simulations of the early phases of open cluster evolution
including a large population of planetesimals, initially arranged in Kuiper-belt like discs around each star. Using a new, 4th-order and time-reversible $N$-body code on Graphics Processing Units (GPUs), we evolve the whole system under the stellar gravity, i.e. treating planetesimals as test particles, and consider two types of initial cluster models, similar to IC348 and the Hyades, respectively. In both cases, planetesimals can be dynamically excited, transferred between stars or liberated to become free-floating (such as A/2017 U1 or 'Oumuamua) during the early cluster evolution. We find that planetesimals captured from another star are not necessarily dynamically distinct from those native to a star. After an encounter both native and captured planetesimals can exhibit aligned periastrons, qualitatively similar to that seen in the Solar system and commonly thought to be the signature of Planet 9. We discuss the implications of our results for both our Solar system and exoplanetary systems.
\end{abstract}


\section{Introduction}
The vast majority of stars are born in open clusters, which disperse as a result of gravitational scattering between stars on time-scales $\gtrsim100\,$Myr. This is at least an order of magnitude longer than the inferred lifetime of an average protoplanetary disc, in which planetesimals and planets are expected to form around young stars \citep[see e.g.,][]{Alexander2014}. Protoplanetary discs efficiently damp inclination and eccentricity of objects embedded within them \citep[e.g.,][for a review]{Baruteau2014}, essentially shielding a young planetary system from the harsh dynamical environment of the cluster \citep[see e.g.,][]{Picogna2014}. However, the difference in lifetimes between discs and clusters means there is a period of 10s-100s of Myr during which intra-cluster interactions can shape a young planetary system without the presence of a disc to damp away their effects. This implies that intra-cluster interactions could leave a lasting -- and potentially observable -- imprint on young systems of planets and planetesimals. Indeed, as our own Sun was most likely born in an open cluster, such signatures may have been present in the early Solar system and potentially be still observable today. Given that open clusters gradually expand and slowly disintegrate over time, we expect the majority of close and potentially dynamically significant interactions to occur in the first few Myr after disc dispersal, when the cluster is still relatively dense.

Recent observations of open clusters have already begun to reveal a large and diverse population of planetary systems hosted within these already dynamically complex objects. \cite{Zuckerman2013} used the observation of calcium in white dwarfs within the Hyades open cluster to infer the presence of both debris discs and planets around these stars. More recently, evidence has emerged of transiting planets around other Hyads \citep{Livingston2017}. Evidently, there is a growing need to understand the impact of the cluster environment on planetary systems during both their formation and long term evolution. Studying planets or planetesimals in open clusters is a challenging numerical problem due to the vast array of time-scales and sheer number of objects that must be considered.

\cite{Levison2010} actually already performed a simulation of a full open cluster with each star hosting a disc of comets, with a view to explaining the formation of the Oort cloud. They found that the Oort cloud may indeed be formed largely of comets captured from other stars. However, their initial population of comets around each star consisted only of long period, highly eccentric comets similar to our Kuiper Belt's scattered disc, the premise being that the formation of giant planets would generate a considerable number of such objects. Whilst quite plausible, this study did not reveal anything about the potential fate of more traditional Kuiper belt objects. If our Solar system contains signatures of intra-cluster interactions, these populations may be our best hope of observing them.

A follow-up study by \cite{Brasser2012} again modelled an entire cluster - this time with both comets and giant planets - but only one star in each simulation initially hosted comets. They showed that interactions between a primordial asteroid belt, giant planets and other stars in the cluster can explain the formation of our inner Oort cloud, and explain the orbits of objects like Sedna. In a further follow-up study \citep{Brasser2015}, they added giant planet migration to these models, hoping to differentiate between the formation of the inner Oort cloud and a population of extreme Kuiper-belt objects that are "detached" from Neptune. This study lead them to suggest that objects with perihelia $>45\,$au and semi-major axes $>250\,$au belong to the Oort cloud rather than the Kuiper belt.

Whilst \cite{Levison2010} showed that integrating an entire cluster with planetesimals is possible, the more common approach is to somehow decouple the evolution of the cluster and the individual planetary systems. The advantage of this approach is that it drastically reduces computational effort, yet still enables some statistical predictions regarding planetery or cometary systems in clusters. For instance, \cite{Kobayashi2001} performed numerical simulations of stellar flybys - a star with a disc flies past a second, disc-less star. They found that increased eccentricity and inclination in the outer part of a planetesimal disc could inhibit planet formation by causing collision velocities between planetesimals to be greater than their mutual surface escape velocity. In the inner disc however, the eccentricities and inclinations remain almost untouched and planet formation continues unabated. In this work they considered only flybys that were relatively close to the outer disc, without attempting to draw flyby parameters from cluster simulations.

In a similar vein, \cite{Lestrade2011} considered the possibility of stripping debris discs in open clusters, using a combination of numerical simulations of stellar fly-bys and kinetic theory of clusters to estimate the percentage of planetesimals stripped from stars of various masses during a cluster lifetime. Their results suggest that only clusters with initial stellar number densities $>1000\,\mathrm{pc}^{-3}$ pose a significant threat to debris discs. More recently, \cite{Jilkova2016} performed many numerical experiments of a flyby between two stars: one with a disc and one without, in order to study the population of objects captured by the originally disc-free star. They found this mechanism to be a viable way of forming objects with high eccentricity and inclination on the outskirts of planetary systems.

Similar techniques have been used to study the evolution of planets themselves in clusters. \cite{Pacucci2013} considered the incidence of free-floating, Jupiter-like planets in 5 clusters. They use a sub-grid model to describe the changes in planetary orbital parameters as their open-cluster simulations evolve. \cite{Cai2017, Cai2018} simulate the effect of intra-cluster encounters on planetary systems by modeling the cluster and planetary evolution in separate but linked $N$-body simulations. \cite{Pfalzner2018a} combined the results of $N$-body simulations of the open cluster M44 with the results of $N$-body simulations of two stars and their discs flying past each other to ascertain if stellar interactions could affect the formation locations of young planets. \cite{Parker2017} considered the capture of free floating planets in open clusters. To do this, they assumed an initial population of free-floating planets, without making any direct assumption regarding their origin. These studies concentrated on free-floating planets because of their potential detectability and the possibility that one of these planets was captured by our Sun. However, intra-cluster interactions also naturally lead to the liberation of smaller objects from young planetary systems.

Of particular interest in this context is the recent discovery of 1I/2017 U1 'Oumuamua \citep{Meech2017}, the first detected so-called ``inter-stellar comet''. 'Oumuamua was presumably ejected from some other - presently unknown - young planetary system, and has been making its way through interstellar space ever since. It seems prudent to investigate the possibility that objects such as 'Oumuamua could be liberated from their original host system by stellar cluster interactions. The dynamical properties of this interesting object were constrained with extraordinarily high accuracy during its flyby of the Solar system, and we will discuss the implications of these observations later in this paper. \cite{Martin2018a, Martin2018b} already considered the idea that 'Oumuamua might have been ejected from either a young protoplanetary disc or an exo-Oort cloud, finding the latter scenario to be unlikely and the former to be plausible only if the population from which 'Oumuamua originated was anisotropically distributed.

It is, of course, not just alien planetary systems that provide evidence and motivation for studying planetesimal discs in clusters. As mentioned above, \cite{Levison2010} already investigated Oort cloud formation in the context of open clusters, and our own Kuiper belt displays some very interesting dynamical characteristics which may to some extent be the result of interactions in the Sun's natal environment. Objects within the belt occupy a broad and varied region of phase space, which can be used to break the population down into distinct families. \cite{Elliot2005} report on the Deep Ecliptic Survey (DES) looking for Kuiper Belt Objects (KBOs). They introduced a convenient scheme for separating our Kuiper belt into distinct dynamical populations, in part using the Tisserand parameter of each object:
\begin{equation}\label{eq:tiss}
T_p = \frac{a_p}{a} + 2 \mathrm{cos}(i) \sqrt{(a/a_p)(1-e^2)},
\end{equation}
where $a$ and $a_p$ are the semi-major axes of the KBO and perturbing body respectively, $i$ is the inclination of the KBO relative to the perturber and $e$ is the eccentricity of the KBO orbit. In the case of the DES, the perturbing body is Neptune. They use the Tisserand parameter as well as other dynamical considerations to split their observed KBOs into 5 distinct dynamical categories. 3 of these categories are relevant here:
\begin{itemize}
\item Classical - objects with $e < 0.2$ and $T_p > 3$ relative to Neptune
\item Scattered - objects with $T_p < 3$ relative to Neptune
\item Scattered-extended - objects with $e > 0.2$ and $T_p > 3$ relative to Neptune
\end{itemize}
We will refer to these definitions throughout this manuscript, using the term KBOs to refer to all of these objects together.
 
Here we aim to perform a complete analysis of how Kuiper-style planetesimal belts are affected by a cluster environment, investigating all of the cluster-related effects mentioned above. We perform $N$-body simulations of stars embedded in young clusters where each star hosts its own Kuiper-belt-style disc of planetesimals, and make no attempt to decouple the evolution of the planetesimal populations from the rest of the cluster. While this approach is computationally expensive, it allows us to gather statistics for planetesimals across an entire cluster, and understand the likelihood of various different outcomes for different masses of star. In particular, we are interested in understanding how intra-cluster interactions might have shaped the present day population of our Kuiper belt, and if there is a chance that these populations contain material captured from another star.

\section{Numerical method}

\begin{table*}
\caption{List of models considered in the text, and some ideal and realized properties of these models in the initial conditions. $N_\mathrm{stars}$ and $N_\mathrm{test}$ are the total numbers of stars and test particles in each simulation respectively. $M_{stars}$ is the mass of stars in each simulation, which varies even between simulations with identical numbers of stars due to the random selection of masses from the Kroupa IMF. $b$ is the Plummer sphere scale parameter. $\rho_\mathrm{core,pred}$ volume mass density of stars that would lie within the core radius ($0.64b$) of an ideal, continuous Plummer sphere with the given mass $M_\mathrm{stars}$ and scale parameter $b$. Since our Plummer spheres are discrete, the actual mass within the core radius varies somewhat. $\rho_\mathrm{core,meas}$ is thus the measured mass within the core radius at the beginning of each simulation. $M(R < R_{50\%})$ signifies the mass within $1.3b$, the radius that would contain 50\% of the mass of an ideal, continuous Plummer sphere. Note that the combination of random selection of mass, position and velocity can lead to considerable variation between clusters with otherwise identical parameters.
\label{tab:models}}

\begin{tabular}{|c|c|c|c|c|c|c|c|} 
\hline 
Model name & $N_\mathrm{stars}$ & $M_\mathrm{stars}$ & $N_\mathrm{test}/N_\mathrm{stars}$ & Scale parameter $b$ & $\rho_\mathrm{core,pred}$ $M_\odot/\mathrm{pc}^3$ &  $\rho_\mathrm{core,meas}$ $M_\odot/\mathrm{pc}^3$  & $M(R < R_{50\%})/M_\mathrm{stars}$ \\ 
\hline 
\texttt{IC348\_1} &440&  233.58&  200 & 0.49 & 283.22  & 172.94  & 41.23 \% \\
\texttt{IC348\_2} &440&  241.08&  200 & 0.49 & 292.31  & 284.82  & 42.05 \% \\
\texttt{IC348\_3} &440&  206.21&  200 & 0.49 & 250.03  & 350.38  & 52.63 \% \\
\texttt{IC348\_4} &440&  227.42&  200 & 0.49 & 275.75  & 149.26  & 50.29 \% \\
\texttt{IC348\_5} &440&  205.64&  200 & 0.49 & 249.33  & 226.82  & 52.84 \% \\
\texttt{IC348\_6} &440&  240.47&  200 & 0.49 & 291.57  & 310.13  & 53.90 \% \\
\texttt{IC348\_tight} & 440&  252.37&  200 & 0.49 & 306.00  & 323.86  & 52.11 \% \\
\texttt{Hyades\_b1.2\_1} &  2250&  1163.66&  200 & 1.20 & 96.06  & 88.70  & 47.35 \% \\
\texttt{Hyades\_b1.2\_2} &  2250&  1197.97&  200 & 1.20 & 98.89  & 103.24  & 49.91 \% \\
\texttt{Hyades\_b1.2\_1\_rerun} & 2250&  1123.95&  200 & 1.20 & 92.78  & 65.20  & 45.78 \% \\
\texttt{Hyades\_b1.2\_2\_rerun} & 2250&  1153.03&  200 & 1.20 & 95.18  & 74.11  & 47.12 \% \\
\texttt{Hyades\_b1.2\_tight} & 2250&  1125.69&  200 & 1.20 & 92.93  & 92.89  & 45.96 \% \\
\texttt{Hyades\_b0.6\_1} & 2250&  1137.57&  200 & 0.60 & 751.26  & 704.91  & 47.75 \% \\
\texttt{Hyades\_b0.6\_2} & 2250&  1165.45&  200 & 0.60 & 769.68  & 751.72  & 47.85 \% \\
\hline 
\end{tabular} 
\end{table*}

We use the time-integration algorithm named `4A' by \cite{ChinChen2005} and originally discovered by \cite{Suzuki1995, Chin1997}. Like the popular leapfrog method, this integrator is symplectic and time-reversible, but fourth-order rather than second-order accurate. A single time step takes only three times longer than for the leapfrog but may achieve errors that are orders of magnitude smaller than for the leapfrog. This means that significantly less computational effort must be extended to achieve an equivalent integration accuracy. Given that the life-time of an open cluster is of order 100 Myr and Kuiper Belt objects orbit with a period of order 100 yr, we wish to achieve minimal errors while keeping run-times manageable. The 4A algorithm is ideal for this, and we have found its energy conservation properties to be far superior to Leapfrog even with much longer time-steps. 

Our code uses a global, adaptive time-step applied to all particles. In principle, an individual time-stepping scheme could speed up these simulations somewhat. However, implementing per-particle time-stepping in a time-reversible (and energy-conserving) fashion is challenging \citep[see e.g.,][]{Dehnen2017}, and with our GPU implementation of the 4A algorithm, the $N^2$ force calculation is already very quick. We therefore stick with a global time-step, but ensure that it is adapted in a time-reversible fashion. The full details of this scheme are given in Appendix~\ref{appen:time}.

\section{Initial conditions}

\begin{figure*} 
\includegraphics[width=0.5018\linewidth]{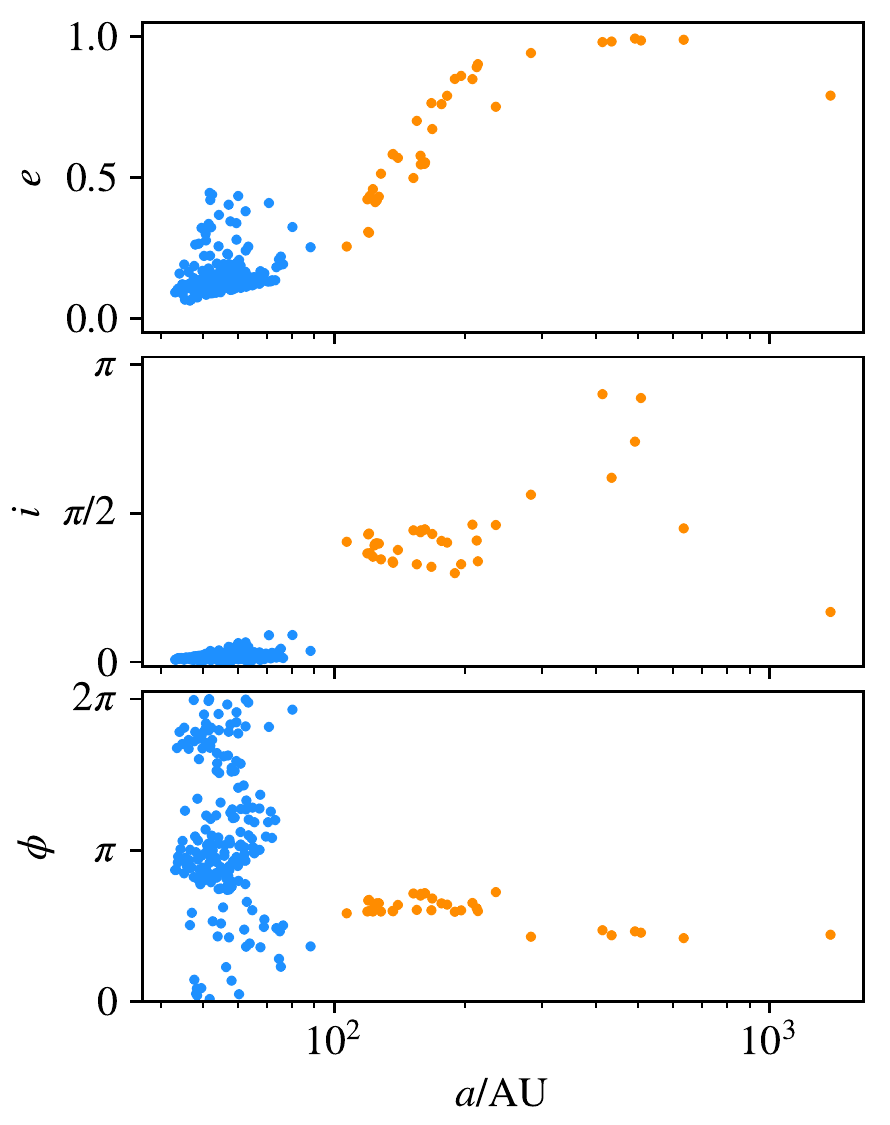}
\includegraphics[width=0.4355\linewidth]{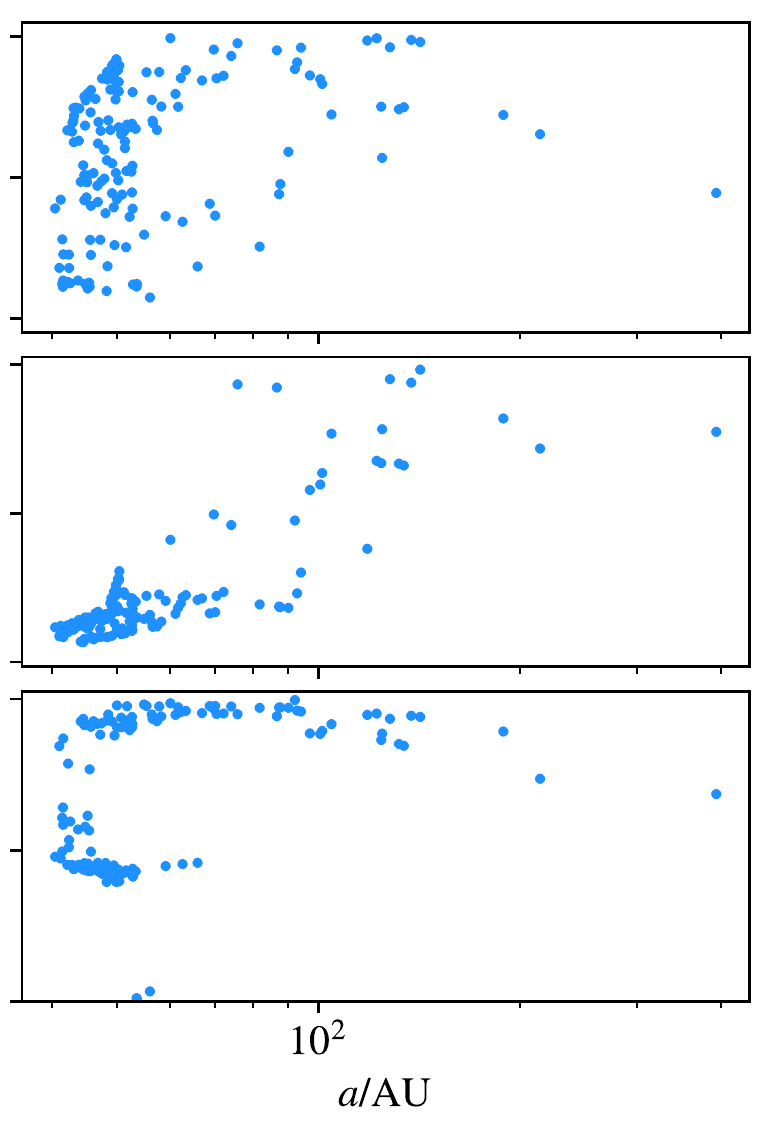}
\caption{Example results from simulation \texttt{IC348\_5} after 10\,Myr of cluster evolution. The left and right panels show the orbital elements (eccentricity $e$, inclination $i$ and longitude $\phi$ of periastron) of planetesimals hosted by a 0.73 $M_\odot$ and 0.13 $M_\odot$ star, respectively. 
The two stars had a mutual encounter that dynamically excited both discs, causing the lower mass star to lose 37 of its primordial test particles. All of these became attached to the more massive star, forming a small and aligned scattered-extended disc. The encounter also scatters a large number (38\%) of the larger star's primordial planetesimals onto orbits that meet the DES definition of ``scattered''. As a consequence, the star is left with a Kuiper belt with 3 distinct populations similar to our own. The less massive is left only with objects in the ``scattered'' regime. Note that in both cases, the primordial planetesimals (blue) exhibit alignment of periastron in their outer regions after the encounter. The pericentre distance of the two stars during the encounter was $\simeq 1045$au, with an eccentricity (excluding the rest of the cluster potential) of $e \simeq 1.12$ \label{fig:pl440_star296} }
\end{figure*}

We base our cluster initial conditions on the standard \cite{Plummer1911} model. Stellar masses in our clusters are drawn from the \cite{Kroupa2001} initial mass function (IMF), truncated below $M_*=0.08 M_\odot$ and above $M_*=20 M_\odot$. This leads to an average stellar mass of $\simeq 0.51 M_\odot$. We construct a standard Plummer sphere with each particle being assigned a mass from the Kroupa IMF, and then assign a disc of 100 or 200 planetesimals to each star. The naming scheme for our simulations gives information regarding the choice of Plummer sphere initial conditions. Names are presented in the format \texttt{CLUSTER\_bXX\_i}, where \texttt{CLUSTER} states which observed cluster the simulation is based on, \texttt{XX} represents the \cite{Plummer1911} scale parameter $b$ in units of parsecs, and \texttt{i} represents which realisation of a particular cluster model is being considered. The individual planetesimal discs are initially dynamically cold, with near-circular orbits and zero mutual inclination, but each individual disc has a random inclination (chosen from a uniform distribution) relative to the other discs in the cluster. They are test particles within our simulations, i.e.\ interact with every star, but not with other planetesimals. 

The test particles are distributed around each star between a randomly selected inner and outer radius, $r_{\mathrm{in}}$ and $r_{\mathrm{out}}$ respectively. We choose $r_{\mathrm{in}}$ from a uniform distribution, for which the lower limit is the radius at which a circular orbit would have some minimal orbital period -- a parameter which varies between simulations -- and the upper limit is always the radius for which the orbital period is $\simeq 715\,$yr (80\,au in the Solar system). For the bulk of our models, the minimum orbital period in the disc is that of Neptune ($\simeq 164.3$yr), or in other words, the inner edge of our own Kuiper Belt. In this way, all our discs are analogs of the Kuiper Belt, but the more massive stars in our simulations host slightly more radially extended discs, because $a\propto M_*^{1/3}$ at constant period. 

The outer edge of the disc, $r_{\mathrm{out}}$, is then chosen from another uniform distribution, whose lower limit is $r_{\mathrm{in}}$ and whose upper limit is the radius corresponding to an orbital period of 1000\,yr (100au in the Solar system). Initial positions are then sampled randomly from the the annulus between $r_{\mathrm{in}}$ and $r_{\mathrm{out}}$ such that the surface density of particles is the same across the entire annulus. Finally, the particles are given the velocity for a circular orbit at their position, though a combination of truncation and round-off error ensures they still have negligible eccentricities with an average of $e \simeq 2.6 \times 10^{-4}$.

This choice of initial conditions is simplistic by necessity: the Kuiper Belt is the only well-studied planetesimal disc, and we therefore have very little data to base our choice of initial conditions on. In an effort to understand the effect that this simplification has on our initial conditions, we also perform several simulations where the interior period of the planetesimal belt around each star is allowed to be as short as $\simeq 31.6$yr - equivalent to an orbital radius of 10\,au in the Solar system. These simulations are designated by the suffix \texttt{tight} in their names.

As our fiducial cluster model, we consider a well-studied and extremely young cluster: IC348. These simulations have names beginning with \texttt{IC348}. \cite{Luhman2016} have performed an extensive census of this young cluster, which is estimated to be between 2 and 6 Myr old and still partially immersed in its parent molecular cloud. This age suggests that many of the stars will soon lose their protoplanetary gas discs, making this the ideal starting point for our simulations of the cluster environment in the immediate aftermath of disc dispersal. \cite{Luhman2016} list 478 known members of the cluster, which they describe as being nearly complete down to $ \gtrsim 0.01 M_\odot$. Based on this number and the fact that we truncate our Kroupa IMF at $M_* = 0.08 M_\odot$, we take the number of stars as 440 for each simulation of IC348. The cluster is approximately 300\,pc from us with a radius of 14', giving a projected radius of around 1.22pc. This in turn gives us a Plummer scale parameter of $b = 0.49$pc, assuming that 80\% of the mass of the Plummer sphere falls within this 1.22pc radius. This gives a good visual fit to the distribution of stars in Figure 1 of \cite{Luhman2016}. We stress, however, that the aim is not to perform realistic simulations of IC348, but use this cluster as a template for which we may study planetesimals in realistic clusters in general. Finally, each star in our \texttt{IC348} simulations begins with 200 primordial planetesimals, giving a total of 88,000 test-particles in each \texttt{IC348} simulation.

\begin{table*}
\caption{Summary of key results from each IC348 simulation. The precise definitions of the quantities tabulated are given in Section~\ref{sec:results}.
\label{tab:results}}

\begin{tabular}{c|c|c|c|c|c|c|c|c|c|c} 
Model (T/Myr) & $N_\mathrm{ecc}$ & $N_\mathrm{align}$ & $N_\mathrm{loser}$ & $N_\mathrm{strip}$& $N_\mathrm{thief}$ & $N_\mathrm{scat}$ & $N_\mathrm{kuiper}$ & $N_\mathrm{0classic}$ &$N_\mathrm{affected}$  &$n_\mathrm{free}$   \\ 
\hline 
\texttt{IC348\_1} (10) &
\makecell{43\\(9.8\%)} & \makecell{59\\(13.4\%)} & \makecell{25\\(5.7\%)} & \makecell{12\\(2.7\%)} & \makecell{13\\(3.0\%)} & \makecell{17\\(3.9\%)} & \makecell{9\\(2.0\%)} & \makecell{22\\(5.0\%)} & \makecell{76\\(17.3\%)} & \makecell{2811\\(3.2\%)} \\
\texttt{IC348\_2} (10) &
\makecell{46\\(10.5\%)} & \makecell{66\\(15.0\%)} & \makecell{23\\(5.2\%)} & \makecell{6\\(1.4\%)} & \makecell{20\\(4.5\%)} & \makecell{21\\(4.8\%)} & \makecell{11\\(2.5\%)} & \makecell{20\\(4.5\%)} & \makecell{81\\(18.4\%)} & \makecell{2255\\(2.6\%)} \\
\texttt{IC348\_2} (15) &
\makecell{53\\(12.0\%)} & \makecell{59\\(13.4\%)} & \makecell{28\\(6.4\%)} & \makecell{12\\(2.7\%)} & \makecell{19\\(4.3\%)} & \makecell{24\\(5.5\%)} & \makecell{11\\(2.5\%)} & \makecell{24\\(5.5\%)} & \makecell{82\\(18.6\%)} & \makecell{3403\\(3.9\%)}\\
\texttt{IC348\_3} (10) &
\makecell{47\\(10.7\%)} & \makecell{80\\(18.2\%)} & \makecell{12\\(2.7\%)} & \makecell{2\\(0.5\%)} & \makecell{11\\(2.5\%)} & \makecell{24\\(5.5\%)} & \makecell{10\\(2.3\%)} & \makecell{9\\(2.0\%)} & \makecell{88\\(20.0\%)} & \makecell{946\\(1.1\%)} \\
\texttt{IC348\_3} (15) &
\makecell{59\\(13.4\%)} & \makecell{77\\(17.5\%)} & \makecell{15\\(3.4\%)} & \makecell{5\\(1.1\%)} & \makecell{12\\(2.7\%)} & \makecell{29\\(6.6\%)} & \makecell{13\\(3.0\%)} & \makecell{13\\(3.0\%)} & \makecell{88\\(20.0\%)} & \makecell{1393\\(1.6\%)} \\
\texttt{IC348\_4} (10) &
\makecell{23\\(5.2\%)} & \makecell{44\\(10.0\%)} & \makecell{8\\(1.8\%)} & \makecell{2\\(0.5\%)} & \makecell{7\\(1.6\%)} & \makecell{14\\(3.2\%)} & \makecell{6\\(1.4\%)} & \makecell{7\\(1.6\%)} & \makecell{48\\(10.9\%)} & \makecell{652\\(0.7\%)} \\
\texttt{IC348\_5} (10) &
\makecell{33\\(7.5\%)} & \makecell{48\\(10.9\%)} & \makecell{15\\(3.4\%)} & \makecell{6\\(1.4\%)} & \makecell{14\\(3.2\%)} & \makecell{17\\(3.9\%)} & \makecell{9\\(2.0\%)} & \makecell{10\\(2.3\%)} & \makecell{65\\(14.8\%)} & \makecell{1495\\(1.7\%)} \\
\texttt{IC348\_6} (10) &
\makecell{48\\(10.9\%)} & \makecell{57\\(13.0\%)} & \makecell{23\\(5.2\%)} & \makecell{9\\(2.0\%)} & \makecell{10\\(2.3\%)} & \makecell{18\\(4.1\%)} & \makecell{11\\(2.5\%)} & \makecell{21\\(4.8\%)} & \makecell{71\\(16.1\%)} & \makecell{2238\\(2.5\%)} \\
\texttt{IC348\_tight} (10) &
\makecell{33\\(7.5\%)} & \makecell{58\\(13.2\%)} & \makecell{7\\(1.6\%)} & \makecell{2\\(0.5\%)} & \makecell{6\\(1.4\%)} & \makecell{12\\(2.7\%)} & \makecell{7\\(1.6\%)} & \makecell{8\\(1.8\%)} & \makecell{66\\(15.0\%)} & \makecell{513\\(0.6\%)} \\
\texttt{IC348\_tight} (15) &
\makecell{46\\(10.5\%)} & \makecell{57\\(13.0\%)} & \makecell{12\\(2.7\%)} & \makecell{4\\(0.9\%)} & \makecell{8\\(1.8\%)} & \makecell{19\\(4.3\%)} & \makecell{11\\(2.5\%)} & \makecell{12\\(2.7\%)} & \makecell{69\\(15.7\%)} & \makecell{1060\\(1.2\%)} \\

\end{tabular} 
\end{table*}

\begin{table*}
\caption{Summary of key results from each Hyades simulation. The precise definitions of the quantities tabulated are given in Section~\ref{sec:results}.
\label{tab:results2}}

\begin{tabular}{c|c|c|c|c|c|c|c|c|c|c} 
Model (T/Myr) & $N_\mathrm{ecc}$ & $N_\mathrm{align}$ & $N_\mathrm{loser}$ & $N_\mathrm{strip}$& $N_\mathrm{thief}$ & $N_\mathrm{scat}$ & $N_\mathrm{kuiper}$ & $N_\mathrm{0classic}$ &$N_\mathrm{affected}$  &$n_\mathrm{free}$   \\ 
\hline
\texttt{Hyades\_b1.2\_1 (4)} &\makecell{17\\(0.8\%)} & \makecell{59\\(2.6\%)} & \makecell{3\\(0.1\%)} & \makecell{0\\(0.0\%)} & \makecell{1\\(0.0\%)} & \makecell{6\\(0.3\%)} & \makecell{4\\(0.2\%)} & \makecell{1\\(0.0\%)} & \makecell{61\\(2.7\%)} & \makecell{18\\(0.0\%)} \\
\texttt{Hyades\_b1.2\_1 (8)} & \makecell{48\\(2.1\%)} & \makecell{86\\(3.8\%)} & \makecell{13\\(0.6\%)} & \makecell{1\\(0.0\%)} & \makecell{11\\(0.5\%)} & \makecell{17\\(0.8\%)} & \makecell{11\\(0.5\%)} & \makecell{11\\(0.5\%)} & \makecell{98\\(4.4\%)} & \makecell{294\\(0.1\%)} \\
\texttt{Hyades\_b1.2\_2 (4)} & \makecell{30\\(1.3\%)} & \makecell{71\\(3.2\%)} & \makecell{10\\(0.4\%)} & \makecell{3\\(0.1\%)} & \makecell{6\\(0.3\%)} & \makecell{15\\(0.7\%)} & \makecell{10\\(0.4\%)} & \makecell{8\\(0.4\%)} & \makecell{84\\(3.7\%)} & \makecell{297\\(0.1\%)} \\
\texttt{Hyades\_b1.2\_2 (8)} & \makecell{55\\(2.4\%)} & \makecell{105\\(4.7\%)} & \makecell{17\\(0.8\%)} & \makecell{6\\(0.3\%)} & \makecell{9\\(0.4\%)} & \makecell{25\\(1.1\%)} & \makecell{17\\(0.8\%)} & \makecell{15\\(0.7\%)} & \makecell{128\\(5.7\%)} & \makecell{526\\(0.2\%)}\\
\texttt{Hyades\_b1.2\_1\_rerun (4)} &\makecell{33\\(1.5\%)} & \makecell{65\\(2.9\%)} & \makecell{11\\(0.5\%)} & \makecell{4\\(0.2\%)} & \makecell{9\\(0.4\%)} & \makecell{12\\(0.5\%)} & \makecell{4\\(0.2\%)} & \makecell{11\\(0.5\%)} & \makecell{76\\(3.4\%)} & \makecell{496\\(0.2\%)}\\
\texttt{Hyades\_b1.2\_1\_rerun (4)} &\makecell{29\\(1.3\%)} & \makecell{91\\(4.0\%)} & \makecell{8\\(0.4\%)} & \makecell{0\\(0.0\%)} & \makecell{7\\(0.3\%)} & \makecell{9\\(0.4\%)} & \makecell{2\\(0.1\%)} & \makecell{7\\(0.3\%)} & \makecell{100\\(4.4\%)} & \makecell{127\\(0.1\%)}\\
\texttt{Hyades\_b1.2\_tight (8)} & \makecell{44\\(2.0\%)} & \makecell{80\\(3.6\%)} & \makecell{14\\(0.6\%)} & \makecell{2\\(0.1\%)} & \makecell{12\\(0.5\%)} & \makecell{15\\(0.7\%)} & \makecell{8\\(0.4\%)} & \makecell{14\\(0.6\%)} & \makecell{98\\(4.4\%)} & \makecell{443\\(0.2\%)} \\

\texttt{Hyades\_b0.6\_1 (4)} & \makecell{160\\(7.1\%)} & \makecell{318\\(14.1\%)} & \makecell{33\\(1.5\%)} & \makecell{6\\(0.3\%)} & \makecell{22\\(1.0\%)} & \makecell{63\\(2.8\%)} & \makecell{25\\(1.1\%)} & \makecell{36\\(1.6\%)} & \makecell{370\\(16.4\%)} & \makecell{922\\(0.4\%)}\\
\texttt{Hyades\_b0.6\_2 (4)} & \makecell{146\\(6.5\%)} & \makecell{359\\(16.0\%)} & \makecell{23\\(1.0\%)} & \makecell{8\\(0.4\%)} & \makecell{17\\(0.8\%)} & \makecell{52\\(2.3\%)} & \makecell{21\\(0.9\%)} & \makecell{23\\(1.0\%)} & \makecell{405\\(18.0\%)} & \makecell{797\\(0.4\%)} \\

\texttt{Hyades\_b0.6\_2 (6)} & \makecell{206\\(9.2\%)} & \makecell{404\\(18.0\%)} & \makecell{32\\(1.4\%)} & \makecell{11\\(0.5\%)} & \makecell{25\\(1.1\%)} & \makecell{76\\(3.4\%)} & \makecell{29\\(1.3\%)} & \makecell{36\\(1.6\%)} & \makecell{477\\(21.2\%)} & \makecell{1110\\(0.5\%)} \\
\end{tabular} 
\end{table*}

To understand the effect star-star interactions on planetesimal discs in a more massive cluster, we consider another extremely well-studied example: the Hyades. Since the Hyades are some 625\,Myr old, they have already lost much of their initial stars, and consequently the rate of close stellar interactions is much lower today than it was at the cluster's birth. We therefore do not wish to use a model of the Hyades today, but rather a model of their state shortly after formation. Fortunately, \cite{Ernst2011} studied this problem in some detail, evolving various different \cite{King1966} and \cite{Plummer1911} models in a galactic potential in an effort to find the initial conditions that best fit the observed Hyades.

Their best-fitting Plummer model with no primordial binaries for the Hyades contained an initial 2250 stars with masses picked from a \cite{Kroupa2001} IMF. \cite{Ernst2011} defined the cluster scale-radius by requiring that 99\% of the cluster mass be within the Jacobi or tidal radius \citep[see e.g.][]{King1962}. Enforcing this condition leads us to a Plummer sphere with $b \simeq 1.2$pc, corresponding to a Jacobi radius of $14.7$pc for their initial location of the Hyades in a Milky Way potential. In addition to our canonical Hyades models with $b = 1.2$ (simulation names beginning \texttt{Hyades}), we run further models with $b = 0.6$ (names beginning \texttt{Hyades\_b0.6}) but the same quantity of stars, to assess the effect of increased density. In all of these models we assign 100 test-particles per star. This is reduced from 200 in the \texttt{IC348} simulations to offset the extra computational effort of integrating 4-5 times as many stars.

We also do not include primordial binaries in these models, because very little is known about planetesimal discs around binary stars, and we want to avoid another level of uncertainty to our initial conditions. We also note that tight binaries could significantly reduce the time-step of our simulations, making simulation times of several Myr implausible. However, binaries may have significant effects on the fate of planetesimal discs, and we intend to study their importance in the future.

Table~\ref{tab:models} shows the parameters chosen for each of the models we consider, including core densities and masses. Note that there is variation between models with identical parameters due to random sampling of the IMF and Plummer sphere. We model 6 different realisations of an IC348-style cluster in an effort to understand the resulting variations in the outcomes, and to allow us to average certain results over multiple models. We integrate each of these models for 10\,Myr. Additionally, we run the simulations \texttt{IC348\_2}, \texttt{IC348\_3}, \texttt{IC348\_tight} for a further 5Myr, such that we can understand how the population of each simulation might evolve if run for longer. Our Hyades models are integrated for shorter time periods -- 8\,Myr and 4\,Myr for the runs with $b = 1.2\,$pc and $b = 0.6\,$pc respectively -- since the larger numbers of stars make them significantly more computationally expensive. However, we also run the model \texttt{Hyades\_b0.6\_2} for an additional 2\,Myr, again in order to investigate how extending the simulations might change the results. 

The initial conditions for our main Hyades simulations contain an error in the sampling of initial disc inclinations: the inclination of each disc was sampled uniformly between $0$ and $\pi$ rather than uniformly in $\mathrm{cos}(i)$ between $-1$ and $1$. This leads to the discs initially being oriented preferentially toward the $z$ axis of each simulation. However, given the random nature of the Plummer sphere itself and the orbits within, we expect this to make little difference to the results. As a test, we ran 2 additional realisations of \texttt{Hyades\_b1.2} to 4Myr with the sampling method corrected and compared them to our main Hyades simulations at $T = 4$Myr. These are denoted in tables \ref{tab:models} and \ref{tab:results2} with the naming convention \texttt{Hyades\_b1.2\_i\_rerun}, and display very little -- if any -- difference to our main models. We are therefore confident that the sampling error has no effect on the results of our longer Hyades runs.

We note that actual clusters drift apart on Myr time-scales, and therefore we only aim to model the short period directly after gas disc dissipation, when the cluster density is at its highest. However, in the case of an old, developed cluster like the Hyades, we can reasonably expect any of the effects seen in our results to be significantly more pronounced. We further note that despite the absence of Galactic tides or a gas potential in our simulations, our cluster models do spread out with time as a result of mass segregation -- more massive stars sink into the potential, forcing less massive ones outward. Each IC348 simulation takes roughly 2-3 weeks on a single NVIDIA K80 GPU, assuming the time-step does not become too small as a result of very eccentric orbits. The Hyades simulations can take up to 3 months - again, depending upon how the orbits of the test particles change.

\section{Results}
\label{sec:results}
Interactions between stars in the cluster lead to a variety of outcomes for the planetesimal discs, that can shape them in unique ways and generate dynamically interesting populations. Determining to which star -- if any -- a given planetesimal is bound is non-trivial in the cluster potential. We thus turn to the tidal force on a star-planetesimal pair to understand if their mutual attraction is greater than the tidal force from the cluster which pulls them apart. The acceleration of planetesimal $i$ due to the gravity from star $j$ is
\begin{equation}
\vec{a}_{ij} = -\frac{G M_j}{|\vec{r}_{ij}|^3} \vec{r}_{ij},
\end{equation}
where $\vec{r}_{ij}\equiv\vec{r}_{i} - \vec{r}_{j}$. The total acceleration of the planetesimal is then
\begin{equation}
\vec{a}_{i}  = \sum_{j\in*}\vec{a}_{ij}.
\end{equation}
The acceleration that planetesimal $i$ feels from the the rest of the cluster -- excluding star $j$ -- is simply
\begin{equation}
\vec{a}_{i\setminus j}  = \vec{a}_{i} - \vec{a}_{ij}.
\end{equation}
The tidal acceleration with which the cluster attempts to separate the pair is then
\begin{equation}
	\vec{a}_{\mathrm{tidal}}  = \vec{a}_{i\setminus j} - \vec{a}_j
	= \vec{a}_i-\vec{a}_j-\vec{a}_{ij}.
\end{equation}
The tidal force may actually be compressive (if $\vec{a}_{\mathrm{tidal}}\cdot\vec{r}_{ij}<0$), but in general pulls the planetesimal-star pair apart.
Here we use a simple criterion based on the modulus of the tidal acceleration compared to that of their mutual attraction and consider a planetesimal unbound if
\begin{equation}
	|\vec{a}_{\mathrm{tidal}}| \ge |\vec{a}_{ij}|.
\end{equation}
Note though, that $|\vec{a}_{\mathrm{tidal}}|<|\vec{a}_{ij}|$ for a given pair does not necessarily imply boundedness. A given planetesimal may be very close to and experience a huge acceleration from a given star whilst still having sufficient velocity to escape its pull completely. Thus, we only consider a planetesimal to be bound to a star if -- in addition to $|\vec{a}_{\mathrm{tidal}}|<|\vec{a}_{ij}|$ -- the eccentricity of the pair $e<1$. Particles that satisfy this condition for more than one star are considered to be unbound from either star -- there is no clear way to establish which star will dominate their future evolution. We note that only a very few particles in our densest simulations of the Hyades fall into this category. We also tried a simpler criterion -- determining if a planetesimal has $e<1$ relative to one and only one star -- but this led to spurious detections of very-long-period planetesimals that happened to share a similar velocity to a given star, but in reality were much closer to other stars. \footnote{This is possible because the eccentricity ignores the cluster contribution and hence overestimates the gravitational influence of the star.}

  Once we have a census of planetesimals bound to each star, we perform some further analysis on each population to establish the following statistics:

\begin{itemize}
\item $N_\mathrm{ecc}$ -- the total number of stars in a given simulation with a planetesimal population that is at least marginally eccentric relative to the population in the initial conditions. Further details regarding how eccentric systems are selected are provided in Section~\ref{sec:Elements}.
\item $N_\mathrm{align}$ -- the total number of stars in a simulation where the planetesimal population shows evidence of apsidal alignment. The method for identifying aligned systems is described in Section~\ref{sec:apsidal}.
\item $N_\mathrm{loser}$ -- the number of stars in a simulation that lost one or more of their original planetesimals.
\item $N_\mathrm{strip}$ -- the total number of stars in a simulation that lost $> 75\%$ of their original planetesimals.
\item $N_\mathrm{thief}$ -- the number of stars in a simulations that gained one or more planetesimals that were originally orbiting a different star.
\item $N_\mathrm{scat}$ -- the number of stars that have one or more planetesimals meeting the DES \citep{Elliot2005} definition of scattered or scattered-extended objects, as defined by equation \eqref{eq:tiss}.
\item $N_\mathrm{kuiper}$ -- the number of stars in a simulation that have at least one planetesimal in each of the DES categories ``classical'', ``scattered'' and ``scattered-extended'' as defined by equation~\eqref{eq:tiss}, and therefore have a Kuiper belt at least passably similar to our own.
\item $N_\mathrm{0classic}$ - the number of stars in a simulation that no longer have any planetesimals meeting the DES definition of classical KBOs. Note that all our planetesimals start in the classical regime. 
\item $N_\mathrm{affected}$ - the number of stars in a simulation that meet one or more of the above criteria. This gives a direct measure of the number of stars that are affected by intra-cluster interactions.
\item $n_\mathrm{free}$ -- the total number of planetesimals in a given simulation that are not considered bound (according to the criterion developed above) to any star. Note that this does not necessarily mean that the planetesimals are unbound from the overall cluster potential, although in general we find that they are.
\end{itemize}

\begin{figure} 
\includegraphics[width=\linewidth]{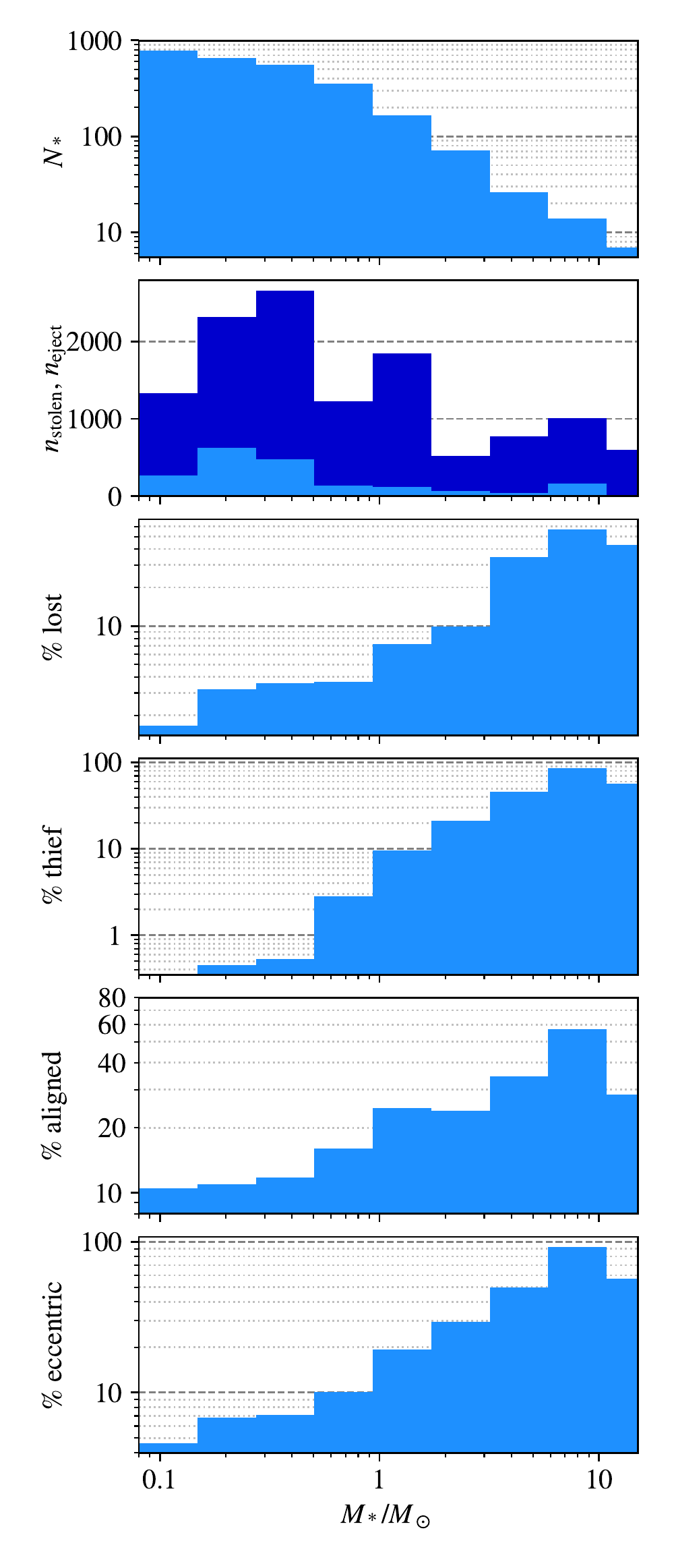}
\caption{Outcomes for planetesimal discs in our 6 simulations of IC348 binned by stellar mass. The top panel shows the total number of stars in each mass bin across all 6 simulations, while the second panel shows the total number of planetesimals stolen from (light blue) or ejected into the cluster and left to float freely (dark blue) in each mass bin. The remaining four panels show the percentage of stars that have lost (3rd panel from top) or stolen (4rd panel) planetesimals, or have populations that exhibit alignment of periapsides (5th panel) or eccentricity growth (6th panel) after 10\,Myr, binned by mass. To avoid low number statistics in the higher mass bins, we stack 6 simulations of IC348 together and use a logarithmic scale in mass.  Discs around higher-mass stars are in general more likely to be affected by the cluster environment. We attribute this trend to multiple effects, discussed in detail in Section~\ref{sec:Elements}. \label{fig:pl440_outcomes}}
\end{figure}

Tables ~\ref{tab:results} and ~\ref{tab:results2} show the final values of these numbers for each of our IC348 and Hyades simulations respectively. Further details regarding the calculation of each of these numbers can be found in Section~\ref{sec:discussion}. Fig.~\ref{fig:pl440_outcomes} shows some of the these values as a function of stellar mass. Figs.~\ref{fig:pl440_star296} and \ref{fig:pl440_star180} show some examples of individual systems that have been sculpted by stellar interactions in our \texttt{IC348} simulations, whilst Fig.~\ref{fig:pl440_tight_all} shows the total population of planetesimals across all stars in one cluster. Note that in figures that display orbital elements, the inclination $i$ and longitude of periastron $\phi$ are calculated using the plane of each star's initial planetesimal disc as the reference plane. In the initial conditions, each star's disc consists entirely of planetesimals whose specific orbital angular momenta $\vec{h}_i$ point in the same direction defined by unit vector $\vec{k}'_*$. The direction of $\vec{k}'_*$ is random and different for each star. We thus calculate inclinations of planetesimals that are bound to a given star relative to that star's $\vec{k}'_*$. The longitude of periapse $\phi$ also requires that we define a reference direction in the plane orthogonal to $\vec{k}'_*$, which we define as $\vec{i}'_* = \vec{k}'_* \times \vec{i}$, with $\vec{i}$ being the unit vector that points along the positive $x$ axis of the simulation. This choice of reference direction is arbitrary but makes little difference to the end result.

\section{Discussion} \label{sec:discussion}

\subsection{General evolution of orbital elements}\label{sec:Elements}
Planetesimals in our simulations start with minimal eccentricity $e \simeq 2.6 \times 10^{-4}$. Therefore, even small levels of dynamical heating in a planetesimal disc should be readily apparent. For the sake of simplicity, we define the population of planetesimals around a star to have experienced eccentricity growth if one or more planetesimals orbiting the star has an eccentricity $e > 0.01$. This criterion is perhaps overly simplistic, but we have found that it effectively identifies systems with interesting features in the eccentricity distribution. For each simulation we count the number of stars that meet this criteria once the simulation is complete. This is the figure $N_\mathrm{ecc}$ in Tables~\ref{tab:results} and ~\ref{tab:results2}. 

\begin{figure} 
\includegraphics[width=\linewidth]{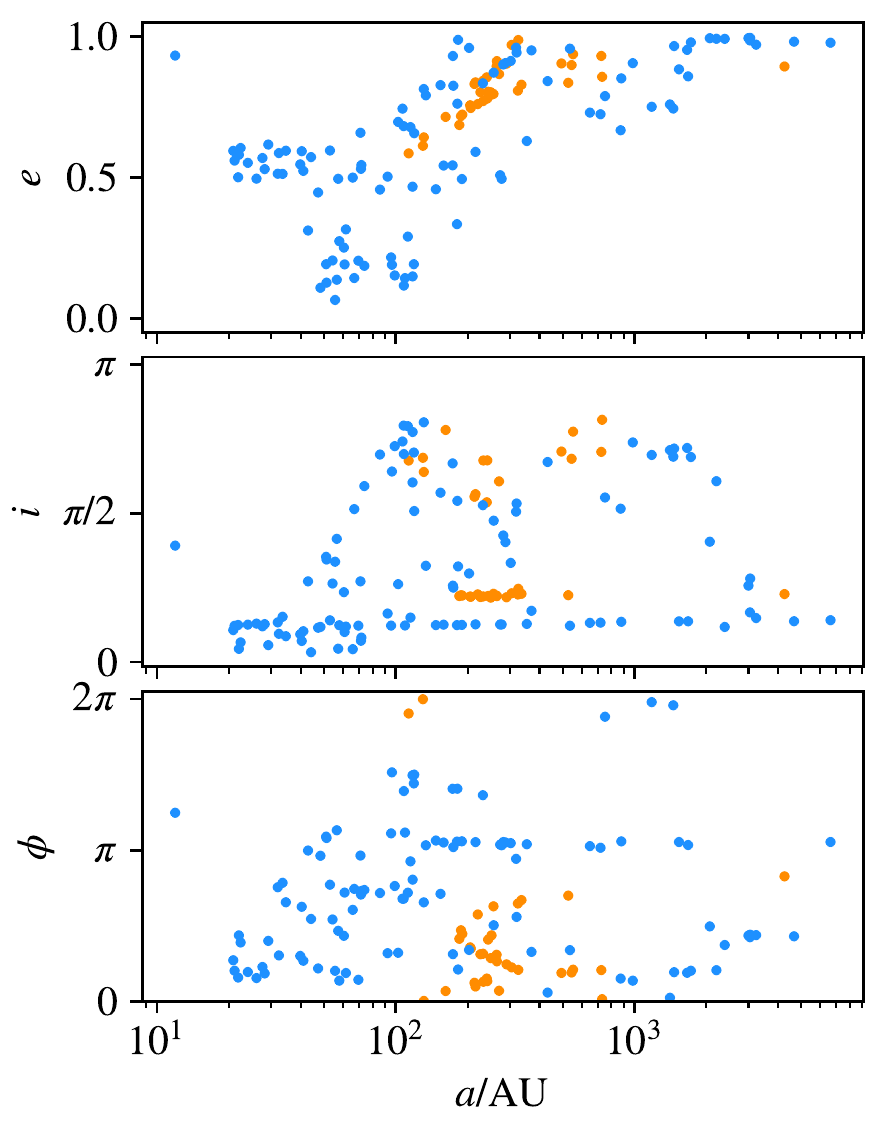}
\caption{An example (taken from the model \texttt{IC348\_2}) of a system around a 1.89 $M_\odot$ star undergoing significant dynamical heating while also capturing a population of planetesimals from another star. The panels from top to bottom show eccentricity $e$, inclination $i$ and longitude of periastron $\phi$. The star had a close encounter with a 1.02 $M_\odot$ star and lost 47\% of its original disc (blue) but captured 18\% of the disc of the smaller star (orange), which in turn lost 86\% of its original disc but captured 16.5\% of the larger star's disc. After the interaction, the captured population around the 1.89 $M_\odot$ star cannot be dynamically distinguished from the primordial planetesimals that were scattered onto high eccentricity, inclined orbits. \label{fig:pl440_6_star65}}
\end{figure}

The numbers in column $N_\mathrm{affected}$ in Table~\ref{tab:results} suggest that up to 20\% of planetesimal discs are affected by the cluster environment in our IC348 simulations. However, the majority of interactions between cluster stars are not violent enough to transfer planetesimals between them, or liberate planetesimals leaving them to float freely. Instead, they lead to an excitation of the planetesimal orbits. This is readily demonstrated by Fig.~\ref{fig:pl440_outcomes} and Table~\ref{tab:results} -- the percentage of planetesimal discs experiencing eccentricity growth is around an order of magnitude more than those experiencing more violent outcomes. On average, for our models of IC348, we find that approximately 10\% of systems undergo interactions within 10\,Myr that generate significant eccentricity. The results of two such interactions can be seen in Figs.~\ref{fig:pl440_star296} and \ref{fig:pl440_6_star65}. As is to be expected, a much smaller percentage of discs demonstrate such dynamical heating in our \texttt{Hyades} models, since their core density is significantly lower. Our \texttt{Hyades} models with $b=0.6$\,pc (and 8 times higher density) do however demonstrate a level of dynamical heating similar to \texttt{IC348} after just 4\,Myr. 

Interestingly, Fig.~\ref{fig:pl440_outcomes} suggests that systems around massive stars are more likely to undergo dynamical heating than those around lower mass stars. In fact, the percentage of stars undergoing dynamical heating at different masses appears to follow a power-law distribution, though further work is required to establish if this relationship holds up with different initial conditions. We first note, that this cannot be explained by the fact that more massive stars have more extended discs, since our scaling to similar orbital periods implies that the binding energy and gravitational pull of the outermost planetesimals scale like $M_*^{2/3}$ and $M_*^{1/3}$, respectively:
it is somewhater harder to affect discs around more massive stars in our simulations.

However, mass segregation dictates that more massive stars sink to the centre of the cluster, where they experience the highest density of stars and are much more likely to experience close encounters. Furthermore, gravitational focussing is stronger for massive stars, increasing the number and severity of close encounters that they experience.

Fig.~\ref{fig:pl440_tight_all} demonstrates how various populations of planetesimals undergo dynamical heating across an entire cluster, and additionally shows the distributions of eccentricity and inclination for various populations. Planetesimals that are ejected by one star and later recaptured by another are likely to end up on Oort cloud style orbits, in a similar fashion to that predicted by \cite{Levison2010}. Interior to the Oort cloud, planetesimals with eccentric or inclined orbits can be either primordial to their star, or captured directly from another star. These planetesimals have orbital elements very similar to those in our own Solar system, with the tail of objects exhibiting long period orbits and high eccentricities in our system being particularly well fit by the simulation results. Note that based on these results, it is not necessarily possible to distinguish between scattered primordial planetesimals and those captured from another star based on orbital elements alone, something demonstrated by the interaction in Fig.~\ref{fig:pl440_6_star65}. Very few planetesimals that are primordial to their star or directly captured from another star end up on Oort-cloud-like orbits, though we note that this might change if additional perturbers such as planets were added to the simulations.

\subsection{Apsidal alignment} \label{sec:apsidal}

\begin{figure}
\includegraphics[width=\linewidth]{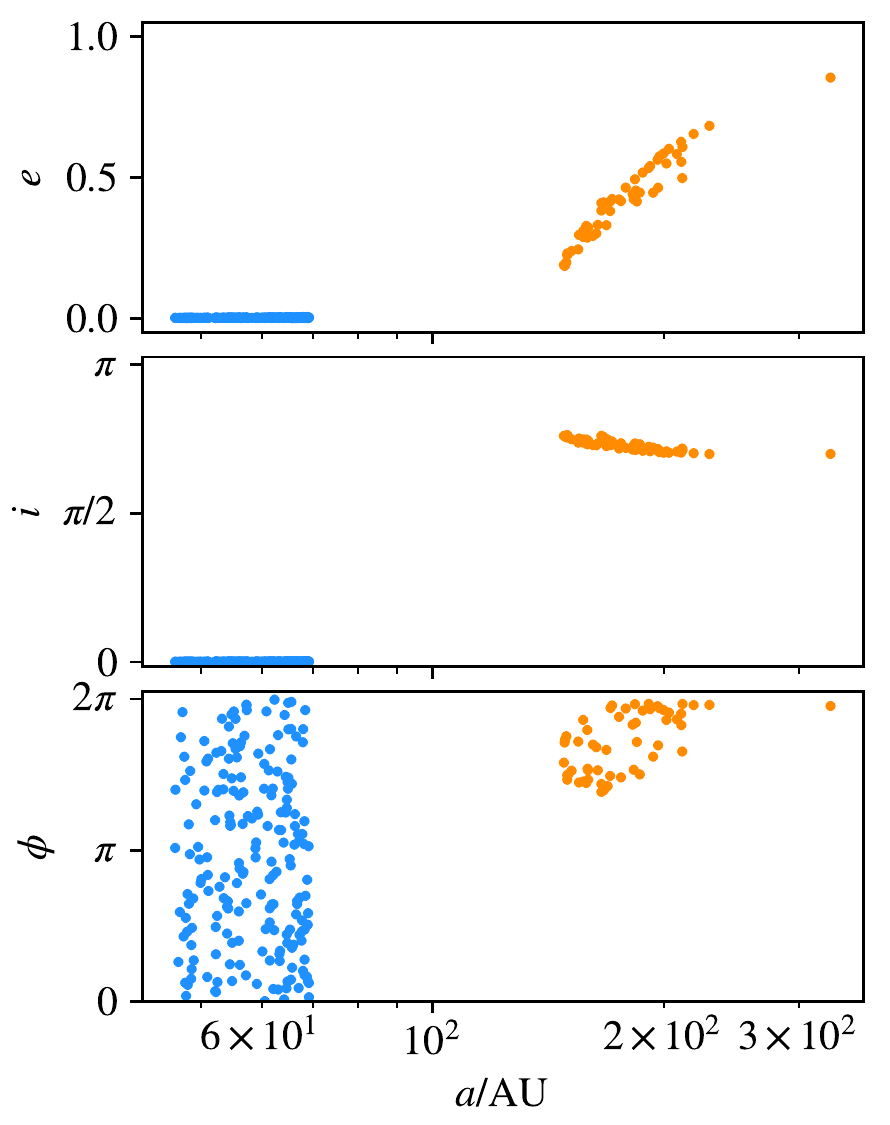}
\caption{An example of apsidal alignment in captured objects from the model \texttt{IC348\_3} after 10Myr. The panels from top to bottom show eccentricity $e$, inclination $i$ and longitude of periastron $\phi$. The colours of the points represent the parent stars of the planetesimals. This 2.4 $M_\odot$ star had a close encounter with a 0.11 $M_\odot$ star, stripping the smaller star of 55\% of its disc and capturing some 50\% of the stripped planetesimals in the process. This captured population (orange) forms a scattered/scattered-extended disc around the star, with the periapsides of the orbits being aligned. The primordial disc (blue) around the star remains largely unaligned, despite clear eccentricity growth. \label{fig:pl440_star180}}
\end{figure}

\begin{figure*}
\includegraphics[width=0.99\linewidth]{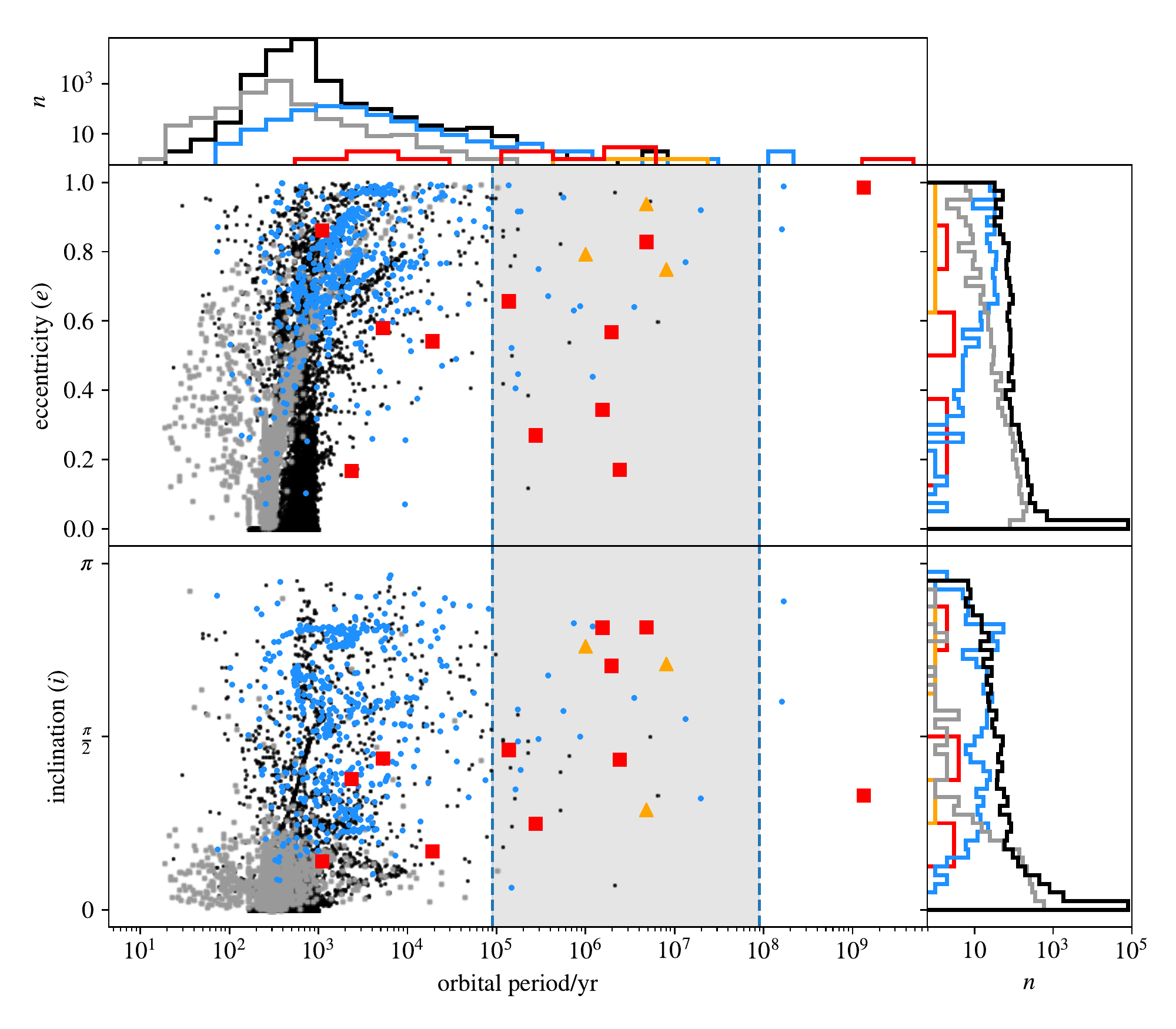}

\caption{Distribution of orbital elements of all bound planetesimals in the model \texttt{IC348\_2} after 15\,Myr. Black points represent planetesimals that remain bound to their original host star. Blue points are planetesimals that have switched their host. The red objects have switched host already after 10\,Myr and at 15\,Myr had switched again, i.e.\ have belonged to at least three different host stars over the course of the simulation. Orange points represent planetesimals that were free floating (bound to no star) at 10\,Myr but have since been re-captured. Finally, grey points represent known trans-Neptunian objects in our Solar system, sourced from the IAU Minor Planet Center. The grey shaded region shows the approximate range of orbital periods in the Oort cloud. \label{fig:pl440_tight_all}}
\end{figure*}

Alignment between the peri-/apo-centres of neighbouring planetesimals is an interesting phenomenon, not least because it is observed in our own system. In exosolar systems, the effect might significantly influence the formation of planets. It may also be observable; an eccentric  debris disc viewed face-on will appear as an extended circular disc if the orbits are randomly aligned, but take the shape of an ellipse if the orbits are similarly aligned.

To look for evidence of apsidal alignment, we consider direction of the eccentricity vector of each planetesimal relative to its current host star. As the eccentricity vector of a planetesimal describes the location at which its periastron occurs, an over-abundance of eccentricity vectors pointing in one particular direction implies alignment within a system. We thus search for planetesimals whose eccentricity vectors point in similar directions - and therefore ``neighbour'' one another. For each pair of planetesimals $1$ and $2$, we compute the angle between their eccentricity vectors as
\begin{equation}
\theta_{1,2} = \mathrm{arccos}\left( \frac{\vec{e}_1 \cdot \vec{e}_2}{|\vec{e}_1||\vec{e}_2|} \right)
\end{equation}
and consider two particles to be neighbours if $\theta_{1,2} \leq \theta_\mathrm{crit}$, with  $\theta_\mathrm{crit}$ being a critical angle. At the beginning of each of our simulations, the eccentricity vectors around each star are randomly distributed in a circular, planar fashion. On average then, we expect each of the $n$ planetesimals orbiting a star to have two ``neighbours'' with eccentricity vectors separated by $\theta_{1,2} \leq 2 \pi/(n-1)$. If we consider larger angular separations by introducing an arbitrary factor $\alpha$ such that the criterion becomes $\theta_{1,2} \leq  \theta_\mathrm{crit} = 2 \alpha \pi/(n-1)$, then the average particle should have $\mu = 2 \alpha$ neighbours in our initial conditions. Assuming the number of neighbours for each particle is distributed in a Poissonian fashion, the standard deviation is $\sigma = \sqrt{\mu}$. If a particle has a number of neighbours that is higher than $\mu$ by a few multiples of $\sigma$, this is strong evidence that it has become a member of an aligned population. We thus consider a system to demonstrate apsidal alignment if it contains one or more planetesimals with a neighbour number $n_b \geq \mu + 4\sigma$, using $\alpha = 5$. The value of $\alpha$ was tuned manually to exclude spurious detections of alignment.


Table~\ref{tab:results} shows that apsidal alignment is a relatively common occurrence in our simulations, occurring for $> 10\%$ of systems in our IC348 simulations, and generally being slightly more common than eccentricity growth. The percentage of systems undergoing apsidal alignment is again lower in our \texttt{Hyades} models than our IC348 models, and again roughly consistent with the number of systems experiencing eccentricity growth. If future observations reveal debris discs that appear elliptical in regions of high stellar density, there is a chance they were shaped by a stellar flyby.

If one considers forming planets in the planetesimal discs left after gas disc dispersion, one might naturally assume that those discs that have undergone dynamical heating in the cluster environment are less likely to bear rocky planets. However, \cite{Kobayashi2001} note that apsidal alignment between neighbouring planetesimals naturally reduces the collision velocity should they meet, thereby reducing the chance that an impact between two planetesimals might destroy them. This in turn means that in the weakly-perturbed inner regions of discs that have undergone stellar encounters, planet formation via the core accretion scenario may not be inhibited. However, the alignment may break down further away from the host star, increasing relative velocities and thereby inhibiting planet formation. This leads to a natural critical radius around stars which have undergone fly-bys, beyond which no planet formation is likely to occur.

\subsection{Capture of planetesimals from other stars}
A few percent of stars in each simulation capture one or more planetesimals that originally belonged to other stars.  The way in which this happens varies greatly between stellar encounters, with some stars capturing planetesimals in a way that makes them indistinguishable from the native population, and others capturing one or two planetesimals on highly inclined and eccentric long-period orbits. Fig.~\ref{fig:pl440_star180} provides an excellent example of a system with a dynamically distinct population: the large population captured from a 0.11 $M_\odot$ star demonstrates orbital elements that are entirely distinct from the primordial disc.

Again, Fig.~\ref{fig:pl440_outcomes} suggests that more massive stars are more likely to thieve planetesimals from other stars. We attribute this to the larger chance of close encounters (due to mass segregation and gravitational focussing) already held responsible in Section~\ref{sec:Elements} for the increased chance of disc heating. As before, the chance of theft also appears to follow a power-law distribution. Massive stars are also more likely to lose some portion of the original disc, but note that this does not necessarily indicate that they are responsible for the majority of captured planetesimals. Indeed, Fig.~\ref{fig:pl440_outcomes} suggests that the majority of captured planetesimals are sourced from the lowest mass stars, simply because there are many more of them.

The number of stars hosting stolen planetesimals is always lower than the number of stars that lost the planetesimals, implying that some small proportion of interactions leads exclusively to the liberation of planetesimals and not their capture.
We note that capture can be a transient phenomenon: planetesimals are often temporarily captured on orbits of $a = 10^5 - 10^6$au and then stripped later by moderate variations in the potential. There is tentative evidence for this in Fig.~\ref{fig:pl440_tight_all}, where the red points represent planetesimals that have been bound to at least three different stars. This raises the interesting idea that the least tightly bound planetesimals in a system may in fact have belonged to several different stars during the cluster phase, eventually ending up permanently bound to whichever host they happened to orbit when that star left its birth cluster. This is in fact similar to the mechanism that \cite{Levison2010} suggest for Oort cloud formation, an idea that we will discuss in further detail later. 

\subsection{Free-floating planetesimals}
Star-star interactions -- particularly in the dense core of a cluster -- can efficiently liberate planetesimals, leading to a population of free-floating planetesimals. These may have been equipped with enough excess energy, i.e.\ 
\begin{equation}
	\label{eq:E}
	E_i=\tfrac12\vec{v}_i^2-\sum_{j\in *}\frac{GM_j}{|\vec{r}_{ij}|} > 0
\end{equation}
to escape the cluster entirely and float freely through the field. Otherwise, if $E<0$ they are destined to drift within the cluster until they either are re-captured by another star, or suffer a close encounter and acquire sufficient energy to escape.

On average between 1 and 4\% of all the planetesimals in our \texttt{IC348} models are free-floating after 10Myr. The difference in number of free-floating planetesimals in the models \texttt{IC348\_2} and \texttt{IC348\_3} between 10 and 15 Myr is around 50\%, suggesting that all models would produce significantly more of these objects if integrated for longer. The second panel in Fig.~\ref{fig:pl440_outcomes} shows that the majority of free-floating planetesimals originally belonged to the lower mass stars in our clusters, but the much lower number of massive stars still contribute a disproportionate amount. Fig.~\ref{fig:pl440_tight_all} demonstrates that some of these planetesimals may eventually be captured by other stars, albeit on distant, easily-perturbed orbits.

\begin{figure*}
	\includegraphics[width=0.99\linewidth]{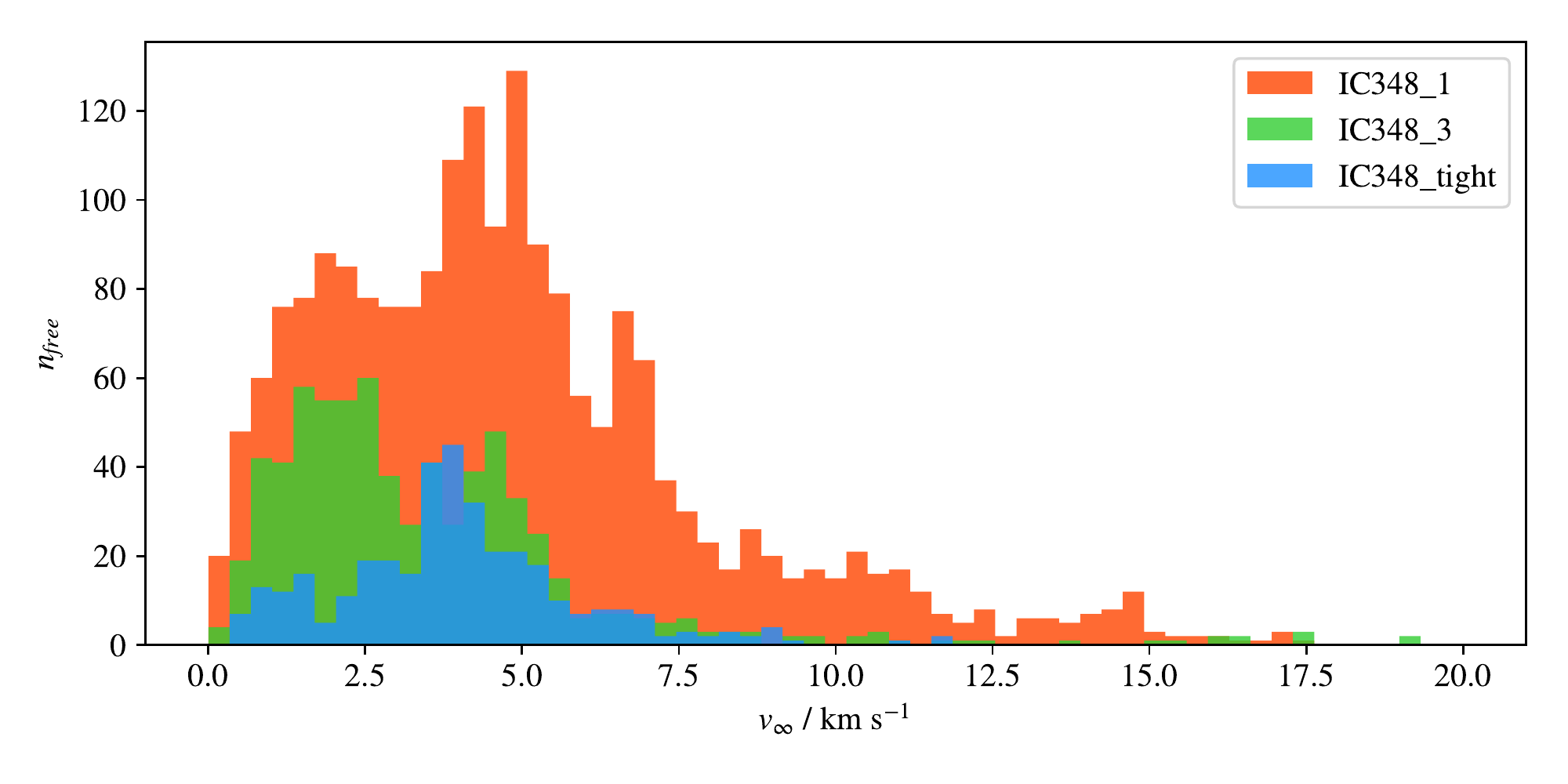}
	\caption{Distribution of hyperbolic excess velocities for planetesimals ejected from our IC348 models after 10\,Myr. The distributions for \texttt{IC348\_2}, \texttt{IC348\_4}, \texttt{IC348\_5} and \texttt{IC348\_6} are not plotted, but are broadly similar to those shown. \texttt{IC348\_1} is anomalous relative to these other models, demonstrating on average notably higher excess velocities, despite the cluster initially having a notably lower core density.
	\label{fig:pl440_vdisp}}
\end{figure*}

Of course, not all planetesimals remain bound to the cluster, and some fraction will slowly drift away into the galaxy. For each planetesimal that we find to be not bound to any star, we compute the energy~\eqref{eq:E} with respect to the cluster. We find that for models of IC348, between 85 and 100\% of all planetesimals that are ejected from their star have $E>0$, i.e.\ will escape the cluster. Fig.~\ref{fig:pl440_vdisp} shows the distribution of hyperbolic excess velocities $v_\infty=\sqrt{2E}$ for planetesimals in our \texttt{IC348} models. All the models of IC348 produce similar distributions of $v_\infty$ -- between 0 and 10\,\kms\ -- though the model \texttt{IC348\_tight} shows a strong peak around 4\,\kms, \texttt{IC348\_1} at 5\,\kms, and \texttt{IC348\_3} shows a preference for velocities around 2.5\,\kms. Table~\ref{tab:models} shows that these models all have vastly different core densities at the beginning of the simulations, which may go some way to explaining these discrepancies.

\subsection{A/2017 U1 (Oumuamua) -- a cluster escapee?}
The prevalence of planetesimals escaping the cluster with $v_\infty$ of a few \kms\ could potentially explain the origin of inter-stellar objects such as A/2017 U1 'Oumuamua. 'Oumuamua had a heliocentric velocity of 26.17\,\kms\ \citep{Meech2017}, but only $\simeq$ 10\,\kms\ relative to the Local Standard of Rest (LSR) \citep[see e.g.,][]{Schoenrich2010}. The velocity dispersion of stars near the Sun is between 25 and 40\,\kms\ \citep[see e.g.,][]{Dehnen1998a,Dehnen1998b,Rix2013}. After its encounter with the Sun, 'Oumuamua's velocity is significantly further from the LSR, As \cite{Meech2017} point out, these facts suggest that 'Oumuamua was ejected from its host system relatively recently and locally,
We can use the velocity profiles in Fig.~\ref{fig:pl440_vdisp} to infer if 'Oumuamua might have been born in and ejected from a local cluster. The excess velocity is of order 1-5\,\kms\ for the vast majority of planetesimals. Assuming then, that 'Oumuamua was ejected from an IC348-style cluster not far from the Sun, it would have an velocity relative to the cluster of between 1 and 5\,\kms, well within the 10\,\kms\ velocity that 'Oumuamua exhibits relative to the LSR. The additional 5-9\,\kms\ difference required to explain 'Oumuamua's initial velocity would then easily be explained by the motion of its birth cluster relative to the LSR. This suggests that if 'Oumuamua was ejected from a local cluster, the ejection was recent, such that the object has had little time to undergo further dynamical interactions that would increase its velocity relative to the LSR.

This scenario is similar to that suggested by \cite{Gaidos2017}, who proposed that 'Oumuamua was ejected from a nearby stellar association by a super Earth or Neptune mass planet with very little excess velocity, and from within a couple of au of its original host star. This limit on the initial orbit of 'Oumuamua around its host was derived from the fact that the object was thought to contain no ice, meaning it must have formed within the ice line of its original host. The debate regarding the composition of the object is still ongoing. Based on a peculiar acceleration evident in its trajectory, \cite{Micheli2018} suggest that 'Oumuamua is in fact undergoing cometary outgassing which is altering the trajectory. \cite{Rafikov2018} finds this scenario unlikely, suggesting that the spin generated by outgassing would have caused 'Oumuamua to break apart during its journey through the Solar system, and \cite{Bialy2018} suggest the peculiar acceleration is instead simply driven by solar radiation pressure. If, however, the object is outgassing, this would naturally favour formation outside the ice line, potentially in a region similar to the discs considered here. Furthermore, \cite{Feng2018} performed backwards integrations of the orbit of 'Oumuamua within the potential of the galaxy, finding that it likely came from a member of the ``Local Association'', which includes several well-known stellar clusters such as the Pleiades. Our simulations suggest that 'Oumuamua could have been liberated from one of these clusters by a stellar flyby, and that there should be many more objects from the same cluster making their way through interstellar space with similar velocities.

\subsection{Destruction of systems}
A rare but important outcome of intra-cluster interactions is stars having their discs almost entirely stripped by an encounter. We define stripping as the removal of $> 75\%$ of the original disc. Between $1$ and $2\%$ of stars in our IC348 simulations meet this criterion, whilst closer to 5\% lose one or more planetesimals. This level of destruction could also easily inhibit planet formation by depriving the young planetary system of solids, so finding a star with no debris disc and no long-period planets could be a sign of a strong stellar interaction.

These results present a slightly different picture to those of \cite{Lestrade2011}, though their study used an analytical prescription to compute the number of close encounters which may explain the differences. Our IC348 models have a core number density $ \simeq 500 \mathrm{pc}^{-3}$, and we find a significant level of planetesimal stripping after just 10\,Myr. \cite{Lestrade2011} suggested that such stripping should only be significant for clusters more dense than $> 1000 \mathrm{pc}^{-3}$, and their work accounted for longer, 100\,Myr time-scales. If we allowed our clusters to evolve for ten times longer, we would certainly see an increase in the level of stripping. For example, in just 5\,Myr the number of stars that have their discs stripped doubles in \texttt{IC348\_tight}, though we note that given the low number statistics, it is not possible to draw any solid conclusions regarding the rate of debris-disc stripping.

\cite{Lestrade2011} also note that the mechanism of cluster-based stripping preferentially strips lower-mass stars, which could lead to the observed paucity of debris-discs around low-mass stars. Our results are consistent with this idea in that the majority of stars we see having their discs stripped have masses $< 1 M_{\odot}$. Again however, given the small number of discs that are stripped overall, it is difficult to draw a definite conclusion from our results regarding a trend in mass. Fig.~\ref{fig:pl440_outcomes} strongly suggests that more massive stars are overall more likely to show evidence of cluster interaction, though this trend does not extend to total destruction of the disc.

\subsection{Application to our own Solar system}
In order to understand the implications of our results for our own Solar system, we first compare the planetesimal discs formed in our clusters to our own Kuiper belt. We follow \cite{Elliot2005}, breaking each disc down into three distinct dynamical populations based on equation \eqref{eq:tiss}. This equation requires the orbital elements of a perturbing body. In the case of our own Solar system, this body is Neptune, but we do not include planets in these simulations. We therefore take the inner edge of each planetesimal disc $r_\mathrm{in}$ in the initial conditions of our simulations, and define $a_\mathrm{p} = 0.95r_\mathrm{in}$. We then define $i$ relative to the initial inclination of the planetesimal disc, such that planetesimals with $i=0$ are aligned with this initial plane. By doing this, we implicitly assume the presence of a Neptune-like perturbing body that is co-planar with and near to the inner-edge of our initial planetesimal discs. We realise that having such a planet in close proximity to our discs may alter some of these results, but this assumption allows us to at least estimate the frequency with which intra-cluster interactions form Kuiper-style-discs.

We find (see columns $N_\mathrm{scat}$ and $N_\mathrm{kuiper}$ in Table~\ref{tab:results}) that debris discs with multiple populations similar to our own Kuiper Belt are readily produced in star clusters. Despite all our stars beginning with Kuiper belts that only have classical, circular populations, many end with an additional population that resembles our own scattered belt, and more often than not, the periapsides of these objects are aligned. We also note that the number of stars $N_\mathrm{0classic}$ with no remaining classical KBOs is of order $1\%$. These stars are generally not the same as the stars that have been stripped, and they normally still host sizeable populations of scattered and scattered-extended objects. This suggests that the destruction of classical Kuiper belt populations by intra-cluster interactions is rare, and -- assuming such debris discs are a common by-product of planet formation -- that the majority of stars host such a disc after leaving their birth cluster.

A second approach is to consider the different types of stellar interactions that our Kuiper belt might have experienced. Assuming our own Solar system was at one point part of an open cluster with a density similar to IC348, Fig.~\ref{fig:pl440_outcomes} allows us to estimate the likelihood that it had certain types of interactions during its early lifetime. For instance, between 3 and 10\% of Solar-mass stars in our IC348 simulations capture planetesimals from another star, whilst between 2 and 7\% lose planetesimals in the cluster environment. So, there is a non-negligible chance that a fraction of our Kuiper belt actually formed around another star or was stripped and carried away by another star.

The clustering in periapsis generated by stellar encounters (particularly in Figs.~\ref{fig:pl440_star296} and \ref{fig:pl440_star180}) is very similar to that seen in the scattered objects of our own Kuiper belt reported by \cite{Trujillo2014}, and invoked by \cite{Batygin2016} as potential evidence for a ninth planet in the outer reaches of the Solar system. However, \cite{Trujillo2014} already showed that the mass in the giant planets of the Solar system is enough to randomise the perihelia of scattered belt objects on relatively short time-scales, and therefore the present-day alignment of perihelia in these objects cannot be a result of a Gyr-old cluster interaction. Nevertheless, our results strongly suggest that some portion of the Solar system's present-day scattered and scattered-extended discs may have been generated by a single, strong cluster interaction -- either from objects previously in the classical disc that were excited to high eccentricity orbits, or objects that were captured from a passing star, or a combination of the two mechanisms \citep[see e.g.,][]{Pfalzner2018b}. This scenario is particularly favourable if there is a 9th planet in the outer Solar system that was captured from a passing star, which may also have donated some of its debris disc to the Kuiper belt \citep[see e.g.,][]{Mustill2016}. We also note that a relatively recent stellar encounter with a field star (perhaps within a few Myr) could have generated a similar alignment that is currently degrading as a result of interactions with the giants planets, though there is no known local star that could be responsible for such an interaction.

Further to this, the Kuiper belt hosts a significant population of objects with significant orbital inclination, for instance "Niku" \citep{Chen2016} with its $i = 110^{\circ}$, $e = 0.3$, $a = 36\,$au orbit. \cite{Batygin2016b} explain these inclined objects in the framework of a 9th planet, invoking \cite{Kozai1962} -\cite{Lidov1962} cycles from planet~9 to generate large inclinations in initially long-period KBOs, before inward scattering by Neptune places them on their current orbits. Most recently \cite{Becker2018} report on the discovery of 2015 BP$_{519}$, with an inclination of 54$^{\circ}$, an eccentricity of 0.92, and a semi-major axis of 450\,au. In the planet~9 framework, this object would be in the process of undergoing Kozai-Lidov cycles, and would presumably be scattered into a tighter orbit in the future. Figs.~\ref{fig:pl440_star180} and \ref{fig:pl440_tight_all} demonstrate that highly inclined planetesimals are readily formed from both captured populations and primordial populations that have undergone scattering. In particular, Fig.~\ref{fig:pl440_star180} shows that large populations of inclined objects can be captured from a lower-mass star without disrupting the captor's primordial planetesimal disc. Some of these objects have large eccentricities that may put them on planet crossing orbits, which could explain  shorter period objects such as Niku. We therefore suggest that some portion of the inclined scattered-extended disc in our Solar system may have been captured in an inclined flyby with a less massive star in the Sun's birth cluster.

Finally, in the vein of \cite{Levison2010} and \cite{Brasser2012}, we consider the formation of the Oort cloud from scattered planetesimals. \ref{fig:pl440_tight_all} shows that stellar interactions can -- in extreme cases -- cause the capture of objects on long-period orbits that would place them within the Oort cloud. The figure also demonstrates that a star's primordial planetesimal population can in a few cases be scattered by a passing star onto Oort-like orbits. Finally, it shows that small numbers of free floating planetesimals can be recaptured in these orbits. Presumably then, there is an element of chance around which star a given long-period planetesimal is orbiting when a cluster dissolves, and planetesimals may be passed between several hosts. More simulations with larger particle numbers are clearly required to understand the formation of this population further, since only a tiny fraction of our Kuiper-style objects meet this fate. What is clear however, is that some portion of our Oort cloud could easily have been carried with the Sun out of its natal star cluster, having originally belonged to other stars. What's more, these planetesimals need not necessarily have been placed on long period eccentric orbits around their parent stars by planetary interactions as previously suggested, but could have been stripped from the dynamically cold inner regions of these stars by violent star-star interactions. This effect presumably accounts for a very small population of the objects in our Oort cloud, and it is not immediately clear how these interlopers might be distinguished from native objects, unless their composition is significantly different from our native planetesimals as to be detected in cometary trails. Future simulations could attempt to investigate the differences in orbital properties between native and non-native Oort cloud objects, but this is beyond the scope of the present study.

\subsection{Caveats \& Future work}
In this work we have investigated the effect of all other stars in an open cluster on planetesimal discs over a long time-scale, but there are some simplifications our simulations that could modify the results described above which we wish to address in future work. To begin, the majority of stars host planets, and including these planets in our simulations could significantly increase the fraction of liberated planetesimals. Planetesimals scattered by star-star interactions might be pushed onto planet-crossing orbits which might then further affect the orbits of the planetesimals. We suggest therefore that our simulations provide only a lower limit on the fraction of free floating planetesimals in a cluster, and that planet-planetesimal interactions may greatly increase this number. It would also certainly increase the number of Oort cloud style objects formed in our simulations. The natural way to explore this problem is simply to introduce planets interior to the debris discs. This, however, would significantly increase the run time of the simulations. There is also naturally the question of how realistic our assumed initial conditions are. It would be prudent in future to try other cluster models such as the \cite{King1966} model or the fractal model favoured by \cite{Parker2017}. Using different initial distributions of planetesimal orbits might also make a difference to the results, though the difference between our \texttt{tight} models with shorter period planetesimals and our standard models is sufficiently small that it seems our conclusions are robust to such changes.

In terms of cluster dynamics, the most obvious omissions here are primordial binaries, the potential of gas left over from star formation, and the galactic potential. We intend to include primordial binaries in a future study, though we anticipate that binary stars would only serve to exacerbate many of the effects seen in this study. In this instance we did not include galactic tides since our integrations last for only around 10\% of an average cluster lifetime and our clusters are still relatively dense, so the potential of the galaxy is unlikely to have a meaningful effect. The potential of leftover gas from cluster formation may, however, have an effect on the time-scales considered here. Young embedded clusters form in molecular clouds, the remnants of which are expelled from the cluster by stellar heating and outflows. This expulsion of gas can cause the cluster to expand and reach a new equilibrium, or disintegrate entirely, depending upon the balance between star formation rate and gas expulsion rate \citep{Lada2003}. 
	
Common practice in cluster simulations is to include a gas potential at the start of a simulation and remove it after a few Myr \citep[for instance, 3\,Myr in][]{Levison2010}, causing the cluster to drift apart more rapidly. This may be particularly relevant to our simulations since IC348 appears to still be partially embedded in the edge of the Perseus molecular cloud \citep[see e.g.][]{Lada1994,Muench2003}. Some observations even suggest that the most gas-rich regions to the southwest of the cluster's core are still forming stars, though the core itself appears to be devoid of star formation \citep{Muench2003,Muench2007}. \cite{Muench2007} show that the core of the cluster where the majority of class II sources are hosted is potentially even still embedded in a filament of gas. Assuming that these class II sources represent stars with gaseous protoplanetary discs, it may still be several million years before they lose their gas. The mutual evolution of the class II sources and the filament in which they reside during this time may have a profound effect on stellar density and therefore the rate of close stellar interactions at the critical point when the planetesimals lose the protection of a gas disc. The exact effect of the removal of this filament is impossible to predict without knowing the exact mass and spatial extent of the gas, making it difficult to simulate. Furthermore, since the cluster is only partially embedded in the cloud, the removal of gas is less likely to be totally catastrophic and might not reduce the density of the cluster by a large factor. We also note that many of the effects observed in our models could impact the younger protoplanetary discs, even in the gas-rich phase. For instance, whilst changes to orbital elements and apsidal alignment would be damped by the gas disc, theft and capture of planetesimals would still be possible. 

Finally, it is important to consider the effect of the total simulation times on the results. We restricted them to $\lesssim$ 10\,Myr in an effort to understand the effect of the cluster environment at its densest point whilst limiting the computational demands of our simulations. Indeed, \cite{Muench2007} calculate the crossing time for IC348 to be at most 1.2 Myr and the relaxation time of the cluster to be roughly $5\times$ this, meaning our simulations ran for almost two relaxation times. As the cluster expands relatively rapidly during this time, we expect the rate of meaningful stellar encounters to drop off quickly. However, for some systems these less frequent encounters might still have a significant effect. Furthermore, \cite{Muench2007} also show that IC348 is already dynamically relaxed and mass segregated, suggesting the mass segregation seen in our simulations would most likely occur when the protoplanetary discs still contain enough gas to damp the dynamical effects described here.
Perhaps then, in future work it would make sense to relax the initial conditions first, lest the mass segregation process have an effect on the results.

By allowing \texttt{IC348\_2}, \texttt{IC348\_3} and \texttt{IC348\_tight} to run for an extra 5\,Myr in addition to the 10\,Myr we ran most of our \texttt{IC348} models for, we are able to gain some insight into how longer simulation times might affect our results. Table~\ref{tab:results} suggests that the extra 50\% simulation time does indeed not translate to an extra 50\% occurrence in all of our outcomes. For instance, the number of stars with partially eccentric discs increases only by around a third, and the number of stars that have lost planetesimals to the cluster increases by an even smaller fraction. Clearly the rate of close encounters has slowed, but there is still some significant evolution. This shows that intra-cluster interactions continue to be important for planetesimal discs even after a couple of relaxation times, though the removal of a gas potential and subsequent expansion of the cluster could change this picture. We also note that there is a general trend for the number of aligned systems to decrease slightly over the extra 5Myr integration time, suggesting that the cluster potential might destroy alignment in some fraction of systems on a relatively short time-scale. Fig.~\ref{fig:pl440_tight_all} also demonstrates that on longer time-scales, other effects may become interesting, such as planetesimals being recaptured from the cluster or passed from star to star on Oort-cloud style orbits. We also ran the model \texttt{Hyades\_b0.6\_2} for an additional 2\,Myr on top of its original 4\,Myr run-time, though the changes to the results in this case are slightly less pronounced than those in \texttt{IC348\_2} and \texttt{IC348\_3}.  This is to be expected since our \texttt{Hyades\_b0.6} have a high core density that shortens the crossing and therefore relaxation time-scales relative to our IC348 models. There is however still some change in planetesimal dynamics over this extra 2\,Myr, hinting that longer simulations may help to further elucidate how Kuiper belts and Oort clouds respond to a cluster environment. This further implies that our default \texttt{Hyades} models - which have the lowest core densities and therefore longest crossing times of all - may continue to evolve at their slow pace for quite some time after the 8\,Myr considered here.

The clear solution to most of the issues and omissions discussed here is to improve the efficiency of our integration method. This will allow for longer runtimes and shorter-period orbits, which enable the addition of stellar binaries and planets and also allow us to explore more massive or denser clusters. We plan to investigate the additional effects listed above in follow-up studies.

\section{Summary}
We have presented $N$-body simulations of planetesimal discs analogous to our own Kuiper-belt, orbiting stars in open clusters. Close interactions between stars in the cluster have a variety of effects on the planetesimal discs, ranging from minor dynamical heating to total destruction. The implications of these intra-cluster interactions are far-reaching and applicable to many areas of planetary astrophysics. In the following we briefly summarise the more important findings of these simulations:
\begin{itemize}
\item Between 10\% and 20\% of debris discs in an IC348-style cluster will undergo dynamical heating due to stellar encounters. This heating can generate populations that are qualitatively similar to those in our own Solar system.
\item Apsidal alignment among planetesimals is readily created by stellar fly-bys. This alignment is qualitatively similar to that seen in our own Solar system.
\item Roughly 3-5\% of stars in our IC348 simulations steal planetesimals from other stars. The number of stars losing their native planetesimals is lower, suggesting free-floating planetesimals are often recaptured by other stars.
\item Intra-cluster interactions efficiently liberate planetesimals from interacting stars. The majority of these planetesimals go on to escape the cluster entirely with excess velocities comparable to their initial orbital velocities -- 5\,\kms\ for planetesimals with the orbital period of Neptune.
\item This effect predicts the presence of many interstellar interlopers such as 'Oumuamua.
\item Even in relatively low-density clusters such as IC348, of order 1-2\% of planetesimal discs can be destroyed within the first 10\,Myr of a cluster's lifetime.
\item Higher-mass stars are on average more likely to 
suffer from significant encounters
than their lower-mass siblings.
\item As a consequence, more massive stars are more likely to both steal and lose planetesimals, but lower mass stars are still the source of the majority of free-floating planetesimals.
\item Some of the dynamical properties of Kuiper belt and Oort cloud objects in our own Solar system can be explained by these effects, though it may be difficult to distinguish these from effects caused by other perturbers such as a ninth planet.
\item These effects are naturally dependent upon the core density of the cluster, and stars in a cluster such as the Hyades are much less likely to undergo significant changes due to the cluster environment.

Future studies could include stellar binaries and planets, which should exacerbate many of the aforementioned effects.
\end{itemize}

\section*{Acknowledgements}
This research has made use of data and/or services provided by the International Astronomical Union's Minor Planet Center. The authors would like to thank Simon Grimm and Richard Alexander for useful discussions, and the anonymous reviewer for comments which greatly improved the manuscript.
TOH acknowledges support from the Swiss National Science Foundation grant number 200020\_162930. This work has been carried out in the frame of the National Centre for Competence in Research 'PlanetS' supported by the Swiss National Science Foundation (SNSF).

\bibliographystyle{mnras}
\bibliography{flyby}

\appendix

\section{Time-reversible time-step adaptation algorithm} \label{appen:time}

Following \cite{Dehnen2017}, we define a time-stepping function $T(\vec{p},\vec{q})$ which gives us the ``optimum'' time-step for our simulation. We also define $h$, the actual time-step used by the simulation. In principle, one simply desires that $h = T$. However, things are not quite this simple. Should one wish to use a time-reversible integration method -- thereby avoiding secular energy evolution -- one must ensure that the value of $h$ is also chosen reversibly. There are many potential ways in which one can achieve this - \cite{HolderLeimkuhlerReich1999} suggest treating the timestepping function $T$ as some kind of mean $\mu(x,y)$ between the previous $h_{n-1/2}$ and next $h_{n+1/2}$ time-steps, giving $T(\vec{p}_n,\vec{q}_n) = \mu(h_{n-1/2},h_{n+1/2})$. For a given mean function $\mu(x,y)$, one can then rearrange to find the next time-step explicitly. This method has a well-known issue whereby the value of $h$ begins to ``flip'' on a time-step to time-step basis, between two values bracketing the value of $T$, but often very far away from it. We found that this flipping issue is problematic for almost all of our simulations, because close interactions between stars and their planetesimals can lead to relatively sudden changes of $T$. \cite{HairerSoederlind2005} instead suggest using the derivative of the function $T$ to evolve the value of $h$ as follows:
\begin{equation}
	\label{eq:adapt:h}
	\frac{1}{h_{n+1/2}} - \frac{1}{h_{n-1/2}} = -\frac{\dot{T}_n}{T_n}.
\end{equation}
This scheme is excellent at maintaining $h$ close to $T$, but unfortunately necessitates the calculation of $\dot{T}$. Depending upon the functional form of $T$, this can be an expensive operation, requiring another $N^2$ loop across particles. Here, we leverage the speed with which GPU-based code can perform $N^2$ operations, and choose $T$ in such a way that the overall impact on runtime is minimised.

We begin by defining the dynamical time between two bodies $i$ and $j$ as
\begin{equation} \label{eq:pij}
	P_{ij} = 2 \pi \sqrt{\frac{|\vec{r}_{ij}|^3}{G(m_i + m_j)}},
\end{equation}
where $\vec{r}_{ij} = \vec{r}_{i} - \vec{r}_{j}$. We use this quantity to compute the characteristic, individual time-scale for the interaction between each pair of particles.
These time-scales must be combined to give an optimal time-step for particle $i$ in a way that gives precedent to the shortest time-scales, while still being continuous and differentiable. We therefore choose to define the optimum time-step for particle $i$ as
\begin{equation} \label{eq:taui1}
	\tau_i^{-q} = \sum_{j \neq i} P_{ij}^{-q},
\end{equation}
with integer $q\simeq4$. This form gives precedence to pairs with shorter interaction time scales and approximates $\min_{j}\{P_{ij}\}$, which is the appropriate form if the dynamics of each particle is dominated by one interaction, as is the case for our simulated planetesimals. Since the optimal time step required for the integration of the stellar trajectories is typically much longer than that for the planetesimals, this approach is completely suitable for our simulations. The computation of $P_{ij}$ must evidently be done on a pair-wise basis, and may in principle be done at the same time as calculating gravity with minimal extra computational overhead. However, since we already have to perform a further $N^2$ loop to compute the derivative of $T$ we calculate $P_{ij}$ simultaneously with this derivative.

Having established this per-particle time-stepping criterion, we need to obtain the global time step function $T$ as a continuous and differentiable combination of all the $\tau_i$ that approximates the minimum across all particles. In analogy to the way $\tau_i$ was constructed from $P_{ij}$, one may use
\begin{equation}
	\label{eq:T:m}
	T^{-m} = \sum_{i=0}^N \tau_i^{-m}
\end{equation}
with integer $m>1$. In the absence of close encounters, the shortest time steps $\tau_i$ are those of the innermost planetesimals around each star, which are roughly equal (by construction of our initial conditions). In this case of  $N_\mathrm{short}\simeq N_*$ particles with equal shortest steps $\tau_\mathrm{short}$, the approach~\eqref{eq:T:m} gives $T\simeq\tau_\mathrm{short}/N_\mathrm{short}^{1/m}$, which is too short by a factor $N_\mathrm{short}^{1/m}$. This causes the simulation to take unnecessarily short time-steps, unless large values of $m$ are used, which leads to issues with both maintaining floating-point accuracy and large values of $|\dot{T}|$. One may avoid this specific problem by introducing a factor $N^{-1}$ on the right-hand side of equation~\eqref{eq:T:m}, but then $T$ is $\simeq N^{-1/m}$ too large in the case of $N_\mathrm{short}=1$.

Finally then, we come to a ``two-sum'' approach, whereby the value of $T$ is defined as
\begin{equation} \label{eq:Ttot}
	T = \eta \left( \frac{\sum_{i=0}^N \tau_i^{-n}}{\sum_{i=0}^N \tau_i^{-(n+m)}} \right)^\frac{1}{m}.
\end{equation}
Here we have added $\eta$ as a parameter to control the overall length of the time-step. In the hypothetical situation where a large proportion of particles require an equally small shortest time-step, a factor $N_\mathrm{short}$ appears in both the numerator and denominator of this fraction, thereby cancelling out. We have found that summing the per-particle criteria in this fashion gives an optimum time-step that is very close to the minimum across all particles. The final piece of the puzzle is then to find $\dot{T}$ by differentiating equation \eqref{eq:Ttot}, giving
\begin{multline}
	\frac{\mathrm{d}T}{\mathrm{d}t} =  \frac{\eta }{m}\left( \sum_{i=0}^N \frac{-n}{\tau_i^{(n+1)}} \frac{\mathrm{d}\tau_i}{\mathrm{d}t}\right) \cdot \left( \sum_{i=0}^N \frac{1}{\tau_i^{n}} \right)^{\left( \frac{1}{m} - 1\right)} \cdot \left( \sum_{i=0}^N \frac{1}{\tau_i^{(m+n)}} \right)^{\left(-\frac{1}{m}\right)} \\ - \frac{\eta }{m}\left( \sum_{i=0}^N  \frac{-n + m}{\tau_i^{(n+m+1)}} \frac{\mathrm{d}\tau_i}{\mathrm{d}t}\right) \cdot \left( \sum_{i=0}^N \frac{1}{\tau_i^{(n + m)}} \right)^{\left( -\frac{1}{m} - 1\right)} \cdot \left( \sum_{i=0}^N \frac{1}{\tau_i^{n}} \right)^{\left(\frac{1}{m}\right)}
\end{multline}
where the derivative of each per-particle time-step is given by
\begin{equation}
	\frac{\mathrm{d}\tau_i}{\mathrm{d}t} =  \left( \frac{1}{\sum_{j \neq i} P_{ij}^{-q}}\right)^{\frac{1}{q} + 1}
\cdot \left( \sum_{j \neq i}  \frac{1}{P_{ij}^{q + 1}} \frac{\mathrm{d}P_{ij}}{dt} \right).
\end{equation}
The second bracket here can be computed on a pairwise basis in the same loop as $\tau_i$ as
\begin{equation} \label{eq:brack1}
	\frac{1}{P_{ij}^{q + 1}} \frac{\mathrm{d}P_{ij}}{dt} = \left( \frac{\sqrt{G(m_i+m_j)}}{2 \pi} \right)^{q+1} \frac{3 \pi}{\sqrt{G(m_i+m_j)}} \frac{\vec{r}_{ij} \cdot \vec{v}_{ij}}{|\vec{r}_{ij}|^{\frac{3}{2}(q+1) + \frac{1}{2}}}.
\end{equation}
We can ensure increased numerical efficiency by selecting a value of $q$ which means we need not perform a square-root operation for each pair of particles in order to calculate $\tau_i$ and $\dot{T}$. Here we use $q=4$. Using this value of $q$ means that the $N^2$ loop required to compute the time-step is faster than a standard $N^2$ force loop, since each particle-particle calculation requires only basic arithmetic operations

Some initial testing of this scheme revealed excellent energy conservation and minimal precession of orbits for simple test problems. Unfortunately, in the context of a cluster, we found this scheme presented an additional problem. In these simulations, two stars that are initially well separated can eventually approach each other very quickly. Since the criterion in equation \eqref{eq:pij} takes no account of relative velocities, this approach can happen on a time-scale that is similar in length to the overall time-step. This can in turn lead to large values of $\dot{T}$, causing $h$ to drift away from $T$ and resulting in either inaccurate or slow integration. To combat this effect, we wish to build a term into the time-stepping criterion that can detect close-encounters a few time-steps before they happen. We therefore define an additional pairwise ``velocity-dependent'' time-stepping criterion
\begin{equation}
Q_{ij} = \frac{1}{k} \frac{|\vec{r}_{ij}|}{|\vec{v}_{ij}|}
\end{equation}
where $k$ is a constant that is chosen to control the relative magnitude of $Q_{ij}$ and $P_{ij}$. We also experimented with a second velocity-related criterion of the form
\begin{equation}
Q_{ij} =\frac{|\vec{r}_{ij}|^2}{\vec{r}_{ij} \cdot \vec{v}_{ij}}
\end{equation}
However, we found that the divergence of this criterion to infinity at apo- and peri-centre made it difficult to keep the chosen time-step $h$ close to the value of $T$. The new velocity-dependent criterion needs to be combined with our previous position-dependent criterion equation \eqref{eq:pij} to give the overall optimum time-step for particle $i$. Equation \eqref{eq:taui1} then becomes
\begin{equation}
\tau_i = \left({\sum_{j \neq i} P_{ij}^{-q} +  Q_{ij}^{-q}}\right)^{-\frac{1}{q}}
\end{equation}
with a derivative given by
\begin{equation}
\frac{\mathrm{d}\tau_i}{\mathrm{d}t} = \left( \frac{1}{\sum_{j \neq i} P_{ij}^{-q} +  Q_{ij}^{-q}}\right)^{\left( \frac{1}{q} + 1\right)} \cdot \sum_{j \neq i} \left( \frac{1}{P_{ij}^{q + 1}} \frac{\mathrm{d}P_{ij}}{\mathrm{d}t} + \frac{1}{Q_{ij}^{q + 1}} \frac{\mathrm{d}Q_{ij}}{\mathrm{d}t} \right)
\end{equation}
The term in ${\mathrm{d}P_{ij}}/{\mathrm{d}t}$ is the same as that in equation \eqref{eq:brack1}. The second term can be computed as
\begin{equation}
\frac{1}{Q_{ij}^{q + 1}} \frac{\mathrm{d}Q_{ij}}{\mathrm{d}t} = k^q\frac{|\vec{v}_{ij}|^{q - 1}}{|\vec{r}_{ij}|^{q + 1}} \left( \frac{\vec{r}_{ij} \cdot \vec{v}_{ij}}{|\vec{r}_{ij}|} |\vec{v}_{ij}| - \frac{\vec{v}_{ij} \cdot \vec{a}_{ij}}{|\vec{v}_{ij}|} |\vec{r}_{ij}| \right).
\end{equation}
Again, we use $q=4$ such that we avoid the need for a square root operation during time-step computation. We have found that this scheme leads to faster precession of Keplerian orbits relative to the purely "position-based" approach, though still significantly better than using  constant time-step. Most importantly though, this scheme greatly reduces the deviation between $h$ and $T$ after a fast close encounter between two stars, without requiring us to lower the value of $\eta$ such that the simulation runs more slowly.

The choice of the powers $m$ and $n$, as well as the time-step control parameter $\eta$, is critical in maintaining the accuracy of the integration, whilst keeping the evolution of the time-step stable and minimising simulation runtime. Larger values of $n$ and $m$ bring the overall value of $T$ closer to the minimum time-step across all particles, but also potentially increase the value of the derivative $\dot{T}$. This can -- for sufficiently large $\eta$ -- lead to the selected time-step $h$ drifting far away from $T$. After some testing, we settled on $k=4$, $m=2$, $n=8$ and $\eta=0.015$, finding that this generally keeps the value of $h$ within 10\% of the value of $T$. Additionally, we set limits that require that the selected time-step is within a factor of 5 of the value of $T$. Note that in principle one could simply reset the selected time-step once it drifts too far from $T$. Any time-step on which this happens would then be irreversible, leading to a small secular energy error. In principle as long as these resets do not occur very often, this should not be a problem. However we elected to keep our integration completely \footnote{Note that technically speaking our implementation is \emph{not completely} reversible since the round-off errors inherent to floating-point arithmetic are not either \cite[see e.g.,][]{Rein2018}.} time-reversible by optimising $n$, $m$, $k$ and $\eta$.
 
\section{Code tests} \label{appen:test}

\begin{table*}
\caption{Time evoution of quantities in \texttt{IC348\_5} and  \texttt{IC348\_5\_short} -- the same initial conditions with a shorter time-step. ~\label{sec:testresults} }

\begin{tabular}{c|c|c|c|c|c|c|c|c|c|c} 
Model (T/Myr) & $N_\mathrm{ecc}$ & $N_\mathrm{align}$ & $N_\mathrm{loser}$ & $N_\mathrm{strip}$& $N_\mathrm{thief}$ & $N_\mathrm{scat}$ & $N_\mathrm{kuiper}$ & $N_\mathrm{0classic}$ &$N_\mathrm{affected}$  &$n_\mathrm{free}$   \\ 
\hline 
\texttt{IC348\_5} (2.5) &
\makecell{9\\(2.0\%)} & \makecell{26\\(5.9\%)} & \makecell{1\\(0.2\%)} & \makecell{1\\(0.2\%)} & \makecell{1\\(0.2\%)} & \makecell{5\\(1.1\%)} & \makecell{1\\(0.2\%)} & \makecell{1\\(0.2\%)} & \makecell{27\\(6.1\%)} & \makecell{111\\(0.1\%)} \\
\texttt{IC348\_5\_short} (2.5) &
\makecell{9\\(2.0\%)} & \makecell{22\\(5.0\%)} & \makecell{1\\(0.2\%)} & \makecell{1\\(0.2\%)} & \makecell{1\\(0.2\%)} & \makecell{5\\(1.1\%)} & \makecell{1\\(0.2\%)} & \makecell{1\\(0.2\%)} & \makecell{23\\(5.2\%)} & \makecell{111\\(0.1\%)}\\
\texttt{IC348\_5} (5) &
\makecell{10\\(2.3\%)} & \makecell{37\\(8.4\%)} & \makecell{2\\(0.5\%)} & \makecell{1\\(0.2\%)} & \makecell{2\\(0.5\%)} & \makecell{9\\(2.0\%)} & \makecell{2\\(0.5\%)} & \makecell{1\\(0.2\%)} & \makecell{39\\(8.9\%)} & \makecell{111\\(0.1\%)}\\
\texttt{IC348\_5\_short} (5) &
\makecell{10\\(2.3\%)} & \makecell{40\\(9.1\%)} & \makecell{2\\(0.5\%)} & \makecell{1\\(0.2\%)} & \makecell{2\\(0.5\%)} & \makecell{9\\(2.0\%)} & \makecell{2\\(0.5\%)} & \makecell{1\\(0.2\%)} & \makecell{42\\(9.5\%)} & \makecell{112\\(0.1\%)}\\
\texttt{IC348\_5} (10) &
\makecell{23\\(5.2\%)} & \makecell{44\\(10.0\%)} & \makecell{8\\(1.8\%)} & \makecell{2\\(0.5\%)} & \makecell{7\\(1.6\%)} & \makecell{14\\(3.2\%)} & \makecell{6\\(1.4\%)} & \makecell{7\\(1.6\%)} & \makecell{48\\(10.9\%)} & \makecell{652\\(0.7\%)} \\
\texttt{IC348\_5\_short} (10) &
\makecell{26\\(5.9\%)} & \makecell{41\\(9.3\%)} & \makecell{9\\(2.0\%)} & \makecell{3\\(0.7\%)} & \makecell{9\\(2.0\%)} & \makecell{18\\(4.1\%)} & \makecell{9\\(2.0\%)} & \makecell{5\\(1.1\%)} & \makecell{49\\(11.1\%)} & \makecell{473\\(0.5\%)} \\
\end{tabular} 
\end{table*}

\begin{figure} 
\includegraphics[width=\linewidth]{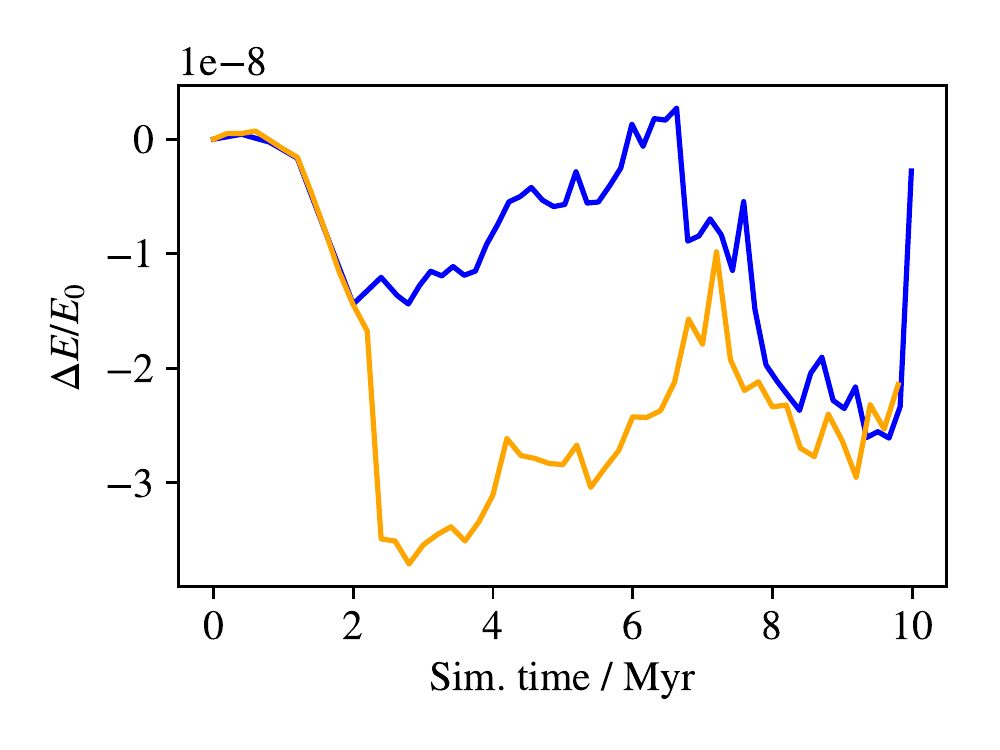}
\caption{Energy conservation over the course of the simulation \texttt{IC348\_5} with $\eta = 0.0075$  (blue) and $\eta = 0.015$ (blue). \label{fig:Energy} }
\end{figure}

Here we demonstrate the efficacy of our new time-stepping method in combination with the 4A integrator, and our new GPU-based implementation thereof. Typically one would do this by showing that the method conserves energy well over the course of an integration. Unfortunately, our planetesimals are massless and therefore do not contribute to the Hamiltonian of the system, so we cannot totally demonstrate the accuracy of the method in this way. As an alternative, we reran the model \texttt{IC348\_5} with $\eta = 0.0075$ rather than 0.015, naming this model \texttt{IC348\_5\_short}. The idea of this test is to show that altering the evolution of the time-step does not significantly impact the results of our simulations, beyond the expected changes due to numerical errors in a chaotic system.

Table \ref{sec:testresults} demonstrates how the results of the simulation change after various integration lengths using the two different values of $\eta$. The differences between the two simulations are minimal and the percentages of star-disc systems undergoing various changes are very similar. Of course, one would not expect such an intrinsically chaotic system to evolve identically with different time-stepping parameters over hundreds of millions of time-steps. We also provide a plot of the total energy error over the course of these two simulations (Fig. \ref{fig:Energy}) to show that the error in the integration of the cluster stars themselves is minimal. The magnitude of the error is similar with both values of $\eta$ since the error is below the level of numerical precision in both cases.

\label{lastpage}

\end{document}